\documentclass[12pt]{article}
\usepackage{amscd,amsthm}
\usepackage{latexsym}
\usepackage{amsmath}
\usepackage{amssymb}
\numberwithin{equation}{section}
\def\bbZ{{\Bbb Z}}
\def\bbR{{\Bbb R}}

\def\Re{{\rm Re}}
\def\Im{{\rm Im}}
\def\const{{\rm const}}
\def\Trace{{\rm Trace}}

\def\a{{\cal A}}

\def\n{{\cal N}}

\def\p{{\cal P}}

\begin{document}
\title{Universality at the Edge of the Spectrum in Wigner Random Matrices}
\author{Alexander Soshnikov\\California Institute of Technology\\Department of
Mathematics\\Sloan 253-37\\Pasadena, CA  91125 USA}
\date{}
\maketitle
\begin{abstract}
We prove universality at the edge for rescaled correlation functions of Wigner random matrices in the limit $n\rightarrow +\infty$.  As a corollary, we show that, after proper rescaling, the 1st, 2nd, 3rd, etc. eigenvalues of Wigner random hermitian (or real symmetric) matrix weakly converge to the distributions established by Tracy and Widom in G.U.E. (G.O.E.) cases.
\end{abstract}

\section{Introduction and Formulation of the \\Results}

We study the classical ensembles of random matrices introduced by Eugene Wigner about forty years ago.  The two models under consideration are Wigner hermitian matrices and Wigner real symmetric matrices.  We start with the hermitian case.

\subsection{Wigner Hermitian Matrices}

The ensemble consists of $n$-dimensional random hermitian matrices $A_n=\Vert a_{ij}\Vert$, where $\Re\ a_{ij}=\Re\ a_{ji}=\xi_{ij}/\sqrt{n},\ \Im\ a_{ij}=-\Im\  a_{ji}=\eta_{ij}/\sqrt{n},\ 1\leq i<j\leq n,\ a_{ii}=\xi_{ii}/\sqrt{n},\ 1\leq i\leq n$, and $\{\xi_{ij},\ \eta_{ij},\ 1\leq i<j\leq n; \xi_{ii},\ 1\leq i\leq n\}$ are independent real variables such that the following conditions hold:

\noindent (i) The random variables $\xi_{ij},\ i\leq j,\ \eta_{ij},\ i<j$, have symmetric laws of distribution,

\noindent (ii) All moments of these random variables are finite; in particular (i) implies that all odd moments vanish,

\noindent (iii)\begin{align}
E(\xi_{ij})^2&=\tfrac{1}{8}=E(\eta_{ij})^2,\ 
1\leq i<j\leq n\\
E(\xi_{ii})^2&\leq \const,\ 
1\leq i\leq n,
\end{align}
with some const $>0$.

We shall denote by const various positive real numbers that do not depend on $n,\ i,\ j$.

\noindent (iv) The distributions of $\xi_{ij},\ \eta_{ij}$ decay at infinity at least as fast as a Gaussian distribution, namely
\begin{eqnarray}
E (\xi_{ij})^{2m},\ \ E(\eta_{ij})^{2m}\leq (\const\cdot m)^m,
\end{eqnarray}

The case of Wigner real symmetric matrices is very similar to the hermitian case, except that we now consider $n$-dimensional real symmetric matrices
$A_n=\Vert a_{ij}\Vert$.

\subsection{Wigner Real Symmetric Matrices}

We assume that $a_{ij}=a_{ji}=\frac{\xi_{ij}}{\sqrt n},\ 1\leq i\leq j\leq n$, and $\{\xi_{ij}\}_{i\leq j}$ are real independent random variables, such that:

\noindent (i) The laws of distributions of $\{\xi_{ij}\}_{i\leq j}$ are symmetric,

\noindent (ii) All moments are finite; in particular, all odd moments vanish,

\noindent (iii)
\begin{align}
E(\xi_{ij})^2&=\tfrac{1}{4},\ 1\leq i<j\leq n,\tag{1.1$'$}\\
E(\xi_{ii})^2&\leq \const,\tag{1.2$'$}
\end{align}

\noindent (iv)
\begin{gather}
E(\xi_{ij})^{2m}\leq (\const\cdot m)^m,\ \  m=1,\ 2,\dots \tag{1.3$'$}
\end{gather}

References [1]--[25], [34], [35], [41], [42] contain an extensive collection of works on the subject from the fifties to the present.  
A famous Wigner semicircle law (see [1], [2], [5]--[12]) can be formulated as follows.  Let
$\lambda^{(n)}_i\geq\lambda^{(n)}_2\geq \cdots \geq \lambda^{(n)}_n$ be the eigenvalues of $A_n$.  The matrix is either real symmetric or hermitian, therefore all eigenvalues are real.  At this point, it is not important which of the two cases we consider.  The eigenvalues of random matrices can be considered as random variables.  One of the fundamental questions is to study their empirical distribution function
$$N_n(\lambda )=\frac{1}{n}\#\{ \lambda^{(n)}_k\leq\lambda ,\ k=1,\ldots ,n\}.$$

The Wigner Semicircle Law claims that $N_n(\lambda )$ converges to a nonrandom limit $N(\lambda )=\int^\lambda_{-\infty}\rho (u)\, du$ as $n\rightarrow +\infty$ where 
$$\rho (u)=\begin{cases}\frac{2}{\pi}\sqrt{1-u^2},& \vert u\vert\leq 1\\
0, & \vert u\vert >1.\end{cases}$$
The convergence is understood to be with probability 1 if entries of all matrices $A_n,\ n=1,2\cdots $, are defined on the same probability space.

The architypical examples of Wigner random matrices are Gaussian Unitary Ensemble (G.U.E.) and Gaussian Orthogonal Ensemble (G.O.E.).  The G.U.E. is the ensemble of random hermitian matrices such that $\Re\ a_{ij},\ \Im\ a_{ij}\break\sim N(0,\frac{1}{8n}),\ 1\leq i<j\leq n,\ a_{ii}\sim N(0,\frac{1}{4n}),\ 1\leq i\leq n$, are independent Gaussian random variables.  Correspondingly, the G.O.E. is the ensemble of random real symmetric matrices such that $a_{ij}\sim N(0, \frac{1+\delta_{ij}}{4n}),\ 1\leq i\leq j\leq n$ are independent Gaussian random variables.  Then the joint distribution of matrix elements can be written as
\begin{equation}
P(dA_n)=\const_{n,\beta}\cdot e^{-\beta \cdot n\cdot \Trace (A^2_n)}dA_n,
\end{equation}
where $\beta =2$ corresponds to the G.U.E., $\beta =1$ corresponds to the G.O.E., and $dA_n$ is Lebesgue measure on $n^2$-dimensional ($\frac{n\cdot (n+1)}{2}$-dimensional) space of matrix elements.  The equation (1.4) implies a nice formula for the induced distribution of the eigenvalues of $A_n$ in the G.U.E. and G.O.E. cases ([17]):
$$dP(\lambda_i,\cdots ,\lambda_n)=P_{n,\beta}(\lambda_1,\cdots \lambda_n)\,
d\lambda_1\dots d\lambda_n,$$
with
\begin{equation}
P_{n,\beta}(\lambda_1\dots\lambda_n)=\const '_{n,\beta}\cdot\sqcap_{1\leq i<j\leq n}\vert\lambda_i-\lambda_j\vert^\beta\cdot e^{-\beta n\cdot (\lambda^2_1+\cdots +\lambda^2_n)}
\end{equation}
which in turn allows one to calculate $k$-point correlation functions of the eigenvalues.  We recall that $k$-point correlation functions are defined as
\begin{equation}
\rho_{n,\beta ,k}(\lambda_1,\ldots ,\lambda_k)=\frac{n!}{(n-k)!}\int_{R^{n-k}}P_{n,\beta}(\lambda_1\dots\lambda_n)\, d\lambda_{k+1}\ldots d\lambda_n,
\end{equation}
where we integrate out the last $n-k$ variables.  For the precise formulas of $\rho_{n,\beta ,k}$ we refer to [17], Ch. 5, 6 and [28], [29].  $k$-point correlation functions are particularly useful in calculating the moments of the number of eigenvalues in an interval $I\subset\bbR^1$.  Let $\nu_{n,I}$ be the number of eigenvalues in $I: \nu_{n,I}=\#\{\lambda^{(n)}_i:\lambda^{(n)}_i\in I,\ i=1,2,\dots n\}$.  Then the mathematical expectation of $\nu_{n,I}$ is given by the formula
\begin{equation}
E\ \nu_{n,I}=\int_I\rho_{n,\beta ,1}(x)\, dx
\end{equation}
and, in general,
\begin{align}
\begin{split}
E\ &\nu_{n,I}\cdot (\nu_{n,I}-1)\dots (\nu_{n,I}-k+1)\\
&=\int_{I^k}\rho_{n,\beta ,k} (x_1,x_2,\ldots , x_k)\, dx_1\dots dx_k
\end{split}
\end{align}
The integration in (1.8) is over the $k$-dimensional cube $I^k=I\times\cdots\times I$.  To study the most challenging problem in the theory of Wigner matrices---the problem of local distribution of 
eigenvalues---one has to consider rescaled $k$-point correlation functions.  Let $x\in R^1$ and consider an interval $I_n(x)$ around $x$ such that the average number of eigenvalues in $I_n(x)$ is of order of constant.  Since
\begin{equation}
E\ \nu_{n, I_n}=\int_{I_n}\rho_{n,\beta ,1}(\lambda )\, d\lambda =\underline{0}(1)
\end{equation}
should imply diam $(I_n(x))\sim\rho^{-1}_{n,\beta ,1}(x)$, we see that for $x$ from the bulk of the spectrum, the intervals $I_n(x)$ will shrink to the point $x$.  Let us define $I_n(x)$ to be
$$(x+c_1/\rho_{n,\beta ,1}(x),\ x+c_2/\rho_{n,\beta ,1}(x)),$$
where $c_1$ and $c_2$ are some fixed constants.  We see that the problem of local distribution of eigenvalues in the neighborhood of $x$ can be reduced to the problem of studying the distribution of the number of particles in $I_n(x)$ as $n\rightarrow +\infty$.  To calculate the moments of $\nu_{n,I_n}=\#\{\lambda_i\in (x+\frac{c_1}{\rho_{n,\beta ,1}(x)},\ x+\frac{c_2}{\rho_{n,\beta ,1}(x)})\}$, we consider a rescaling 
\begin{equation}
\lambda_i=x+\rho^{-1}_{n, \beta ,s}(x)\cdot y_i,\ i=1,\ldots ,k
\end{equation} 
and
\begin{equation}
R_{n,\beta ,k}(y_1,\ldots ,y_k)=\rho^{-k}_{n,\beta ,1}(x)\cdot\rho_{n,\beta ,k}(\lambda_1,\ldots ,\lambda_k).
\end{equation}
The functions at the l.h.s.~of (1.11) are called rescaled $k$-point correlation functions.  It follows immediately from (1.10) that if $\lambda_i\in I_n(x),\ i=1,\ldots ,k$, then the variables 
$y_1,\ldots ,y_k$ are of order of constant, namely $y_i\in (c_1, c_2),\ i=1,\ldots ,k$.  Moreover, the factorial moments of $\#(x+c_1/\rho_{n,\beta ,1}(x),\ x+c_2/\rho_{n,\beta ,1}(x))$ are equal to
\begin{align}
\begin{split}
E\ &\nu_{n,I_n}\cdot (\nu_{n,I_n}-1)\dots (\nu_{n,I_n}-k+1)\\
&=\int_{[c_1,c_2]^k}R_{n,\beta ,k}(y_1,\ldots ,y_k)\, dy_1\cdot dy_k.
\end{split}
\end{align}
To show that $\nu_{n,I_n}$ converges to a limit in distribution as $n\rightarrow\infty$, one needs to show then that rescaled $k$-point correlation functions have a limit too.  This result has been established in the Gaussian cases (see [17]).  Consider first the hermitian case $\beta =2$ (G.U.E.).  Then uniformly on the compact subsets of $R^k$,
\begin{equation}
\lim_{n\rightarrow\infty} R_{n,2,k}(y_1,\ldots y_k)=R_{2,k}(y_1,\ldots y_k)=
\det (K(y_i,y_j))^k_{i,j=1},
\end{equation}
where $K(y,z)$ is an example of a so-called integrable kernel,
\begin{equation}
K(y,z)=\frac{A(y)\cdot A'(z)-A'(y)\cdot A(z)}{y-z}.
\end{equation}
The amazing fact is that $K$ does not depend on $x$, provided $x$ lies in the bulk of the spectrum, $-1<x<+1$. Then one can show that $A(y)=\frac{\sin\pi y}{\pi}$.  If $x=\pm 1$ the kernel still has the form (1.14), but then 
$A(y)=\a i(\pm y)$.  Here $\a i(y)$ stands for the Airy function which is defined as the solution of the differential equation $f''(y)=y\cdot f(y)$ with the asymptotics $f(y)\sim\frac{1}
{2\cdot\sqrt{\pi}\cdot y^{\frac{1}{4}}}
e^{-\frac{2}{3}y^{\frac{3}{2}}}$ as $y\rightarrow +\infty$.  Looking back at (1.10) one may ask about the order of the spectral density $\rho_{n,2,1}(x)$.  The answer is that $\rho_{n,2,1} (x)=n\cdot\frac{2}{\pi}\sqrt{1-x^2}\cdot (1+\bar o(1))$ uniformly on compact subsets of $(-1, 1)$, and 
$\rho_{n,2,1}(\pm 1)=\underline{0} (n^{\frac{2}{3}})$.

In the case of G.O.E., the limiting formulas are slightly more complicated.  
One can show that
$$R_{1,k}(y_1,\ldots ,y_k):=\lim_{n\rightarrow\infty}R_{n,1,k}(y_1,\ldots ,y_k)$$
can be expressed as a square root of the determinant of a $2k$-dimensional matrix consisting of $2\times 2$ blocks $\xi_1(y_i,y_j),\ 1\leq i,j\leq k$.  To define $\xi_1(y,z)$ we start with the real-valued kernel (1.14) and introduce
\begin{eqnarray*}
DK(y,z)&=&-\frac{d}{dz}K(y,z),\\
JK(y,z)&=&-\int^\infty_yK(t,z)\, dt-\tfrac{1}{2}\text{sgn}(y-z)
\end{eqnarray*}
Then
\begin{eqnarray}
\nonumber
& & \xi_1(y,z)= \\
& & \biggl\lbrack\begin{matrix}K(y,z)+ \frac{1}{2} \*\a i(y) \* 
\int_{-\infty}^{z} \a i(t) dt ,\ &\ - \frac{1}{2} \a i(y)\* \a i(z) + DK(y,z)
\\ JK(y,z) + \frac{1}{2} \* \int_{z}^{y} \a i(u) du + \frac{1}{2} 
\* \int_{z}^{+\infty} \a i(u) du \* \int_{-\infty}^{z} \a i(v) dv,\ 
&\  K(z,y) + \frac{1}{2} \a i(z) \* \int_{-\infty}^{y}
\a i(t) dt \end{matrix}\biggr\rbrack
\end{eqnarray}
and
\begin{equation}
R_{1,k}(y_1,\ldots y_k)=\sqrt{\left ( \det (\xi_1(y_i,y_j))^k_{i,j=1}\right )}.
\end{equation}
In [26], [27], [33] Tracy-Widom, and Forrester studied distributions of the first few largest eigenvalues in G.U.E. and G.O.E. (It may be noted that in [27] a Gaussian ensemble of self-dual quaternion matrices, which corresponds to $\beta =4$ in formula (1.5.) was also studied).]  
It was shown that for any positive integer $k$ the joint distribution function of the first $k$ rescaled eigenvalues $(\lambda^{(n)}_1-1)\cdot 2n^{\frac{2}{3}},\ (\lambda^{(n)}_2-1)\cdot 2n^{\frac{2}{3}},\dots ,(\lambda^{(n)}_k-1)\cdot 2n^{\frac{2}{3}}$ has a limit as $n\rightarrow\infty$:
\begin{equation}
F_{\beta ,k}(s_1,\ldots ,s_k)=\lim_{n\rightarrow\infty}P_\beta \left (\lambda^{(n)}_i\leq 1+\frac{s_i}{2n^{\frac{2}{3}}},\ i=1,\ldots ,k\right ).
\end{equation}
The limiting $k$-dimensional distribution function (which is different in G.U.E. and G.O.E. cases) can be expressed in terms of the solutions of completely integrable P.D.E.  The formulas are the simplest when one considers the maximal eigenvalue $\lambda^{(n)}_1$.  Let $q(s)$ be the solution of the Painl\'eve II differential equation $q''(s)=sq(s)+2q(s)^3$ determined by the asymptotics $q(s)\sim\a _i(s)$ at $s=+\infty$.  Then for the G.U.E. $(\beta =2)$:
\begin{equation}
F_{2,1}(s)=\lim_{n\rightarrow\infty}P\left (\lambda^{(n)}_1\leq 1+\frac{s}{2n^{\frac{2}{3}}} \right )=\exp(-\int^{+\infty}_s(x-s)\cdot q^2(x)\, dx)
\end{equation}
and for the G.O.E. ($\beta =1$):
\begin{align}
\begin{split}
F_{1,1}(s)&=\lim_{n\rightarrow\infty}P\left (\lambda^{(n)}_1\leq 1+\frac{s}{2n^{\frac{2}{3}}} \right )\\
&=\exp\left 
(-\frac{1}{2}\int^{+\infty}_sq(x)+(x-s)q^2(x)\, dx\right )
\end{split}
\end{align}
The great attention paid to the local distribution of the eigenvalues has its origin in a general belief that local statistics in random matrices mimic those of the energy levels of highly excited states of heavy nuclei (see e.g. [4] and [17]).  The computer simulations of random matrices ([3], [17], [41]) show that local fluctuations are always the same in the limit $n\rightarrow\infty$ and determined only by the overall symmetries of the ensemble.  This allowed Mehta to formulate the following conjecture that can be found in the introduction of his book on random matrices ([17], p.~9):
\medskip

\noindent{\bf Conjecture (Universality in Wigner Matrices)}

Let $A$ be $n\times n$ real symmetric (hermitian) Wigner random matrices.  Then in the limit of large $n$ statistical properties of $k$ eigenvalues of $A$ become independent of the probability distribution of $a_{ij}$.  In other words, rescaled $k$-point correlation functions tend for every $k$ to the $k$-point limiting correlation functions of Gaussian Orthogonal (Unitary) Ensemble given in (1.15--1.16) and (1.13--1.14).

The purpose of our paper is to establish a universality conjecture both for hermitian and real symmetric ensembles of Wigner matrices (1.1--1.3), 
(1.1'--1.3') at the edge of the spectrum.  We shall prove that in the limit $n\rightarrow\infty$ the distribution of the first few rescaled eigenvalues is independent of the marginal distributions of matrix elements.

\subsection*{Main Result}

Let us consider a Wigner ensemble of $n\times n$ hermitian or real symmetric matrices ((1.1--1.3) or (1.1'--1.3')).  Fix some arbitrary positive integer $k$ and consider the first $k$ largest eigenvalues of a Wigner matrix: $\lambda^{(n)}_1\geq\lambda^{(n)}_2\geq\cdots\geq\lambda^{(n)}_k$.  Then the joint distribution function of $k$-dimensional random vector with the components $(\lambda^{(n)}_1-1)\cdot 2n^{\frac{2}{3}},\dots ,(\lambda^{(n)}_k-1)\cdot 2n^{\frac{2}{3}}$ has a weak limit as $n\rightarrow\infty$, which coincides with that in the G.U.E. (G.O.E.) case.

\noindent {\it Remarks}  

\begin{enumerate}
\item This is the first rigorous result about universality at the edge of the spectrum.  Recently several groups of mathematicians (Pastur, Shcherbina [44],
Deift, Kriecherbauer, McLaughlin, Venekides, Zhou [47], Bleher, Its [46]; see
also [45] and [39]) established universality at the bulk of the spectrum for certain classes of unitary invariant ensembles of hermitian random matrices, when
\begin{equation}
P(d A_n)=\const_n\cdot e^{-n\cdot \Trace V(A_n)}dA_n
\end{equation}
and $V$, for example, is a polynomial of even degree with a positive leading coefficient.  With the exception of a $V$ quadratic polynomial, which corresponds to G.U.E., matrix elements of $A$ are strongly correlated.

\item There is a recent paper by Johansson [36] that claims the universality of rescaled two-point correlation functions in the bulk of the spectrum for quite a general class of hermitian Wigner matrices.

\item
The main result also holds for the smallest eigenvalues of Wigner matrices.

\item
It should be noted that the limiting distribution (1.18) appeared recently in the paper by Baik, Deift and Johansson [48] as a limit of the rescaled distribution of the length of the longest increasing subsequence of a random permutation from $s_n$ (see also [31]).  In a very recent development, Okounkov ([50]) generalized the Baik--Deift--Johansson result for an arbitrary number of the rows of partitions of $n$.  Among other interesting papers on the subject are [51], [52].  We mention that (1.18) also appeared in the paper by Johansson ([37]) on shape fluctuations in certain random growth models in two dimensions.
\end{enumerate}

The idea of the proof is to study linear statistics of the form
\begin{equation}
\sum_{j=1,2,\dots\ : \lambda_j\geq 0}e^{t\cdot\theta_j},
\end{equation}
where $t>0$ and $\theta_j$ are obtained from the positive eigenvalues $\lambda_i>0$ by rescaling
$$\lambda_j=1+\frac{\theta_j}{2n^{\frac{2}{3}}}.$$
It follows from the semicircle law that
$$\#\{\lambda_j>1-\epsilon\}=n\cdot\frac{2}{\pi}\int^1_{1-\epsilon}\sqrt{1-x^2}\, dx\cdot (1+\bar o(1)) \text{(a.e.)}$$
for $0\leq\epsilon\leq 1$.

In particular,
$$\#\{\lambda_j>0\}=\frac{n}{2}+\bar o(n).$$
The main contribution to the linear statistics $\sum_{j:\lambda_j\geq 0}e^{t\theta_j}$ is due to the eigenvalues at the right edge of the spectrum.  Indeed, the subsum of (1.21) over $0\leq\lambda_j\leq 1-\epsilon$ is negligible:
$$\sum_{j:0\leq\lambda_j\leq 1-\epsilon}e^{t\cdot\theta_j}\leq\#\{0\leq\lambda_j\leq 1-\epsilon\}\cdot e^{-2t\cdot\epsilon\cdot n^{\frac{2}{3}}}\leq n\cdot e^{-2t\epsilon\cdot n^{\frac{2}{3}}}=\bar o(1).$$
To prove the universality of $k$-point correlation functions, it is quite convenient to study their Laplace transform:
\begin{equation}
\begin{split}
\int^\infty_{-\infty}\cdots\int^\infty_{-\infty}&\exp (t_1\cdot y_1+\cdots +t_k\cdot y_k)\cdot R_{n, \beta ,k}(y_1,\ldots y_k) dy_1,\dots dy_k=\\
&=E\sum_{j_1\neq\cdots\neq j_k}e^{t_1\cdot\theta_{j_1}}\cdot e^{t_2\cdot\theta_{j_2}}\cdot\cdots\cdot e^{t_k\cdot\theta_{j_k}},
\end{split}
\end{equation}
where $t_1,t_2,\dots t_k>0$.

We now use the notation (1.11) for an arbitrary Wigner matrix with $\beta =2$ corresponding to the hermitian case, and $\beta =1$ to the real symmetric case.  Strictly speaking, $R_{n,\beta ,k}$ are functions in a usual sense only if the marginal distributions of matrix elements are absolutely continuous with respect to the Lebesgue measure. Generally, $R_{n,\beta ,k}$ are understood to be distributions.  The universality then should imply convergence of the r.h.s.~of (1.22) to the limit
\begin{equation}
\int^\infty_{-\infty}\cdots \int^\infty_{-\infty}\exp (t_1y_1+\cdots +t_ky_k)\cdot R_{\beta ,k}(y_1,\ldots y_k)\, dy_1\cdots dy_k.
\end{equation}
Actually we are going to prove a slightly weaker form of the last property.  The reason for the modification comes from the fact that for large positive $\theta_j$'s the corresponding terms in the r.h.s.~of (1.22) are going to be exponentially large and therefore even the question of boundedness of the mathematical expectations in the limit $n\rightarrow\infty$ is not easy (essentially, we have to control the probabilities of large deviations).  To avoid this problem, we consider truncated sums corresponding to $\theta_j$ not 
greater than $n^{\frac{1}{6}}$.  We define
\begin{equation}
S_{n,k}(t_1,\dots ,t_k)=\sum_{j_1\neq\cdots \neq j_k}e^{t_1\theta_{j_1}}\cdots e^{t_k\theta_{j_k}}.
\end{equation}

We thus throw away terms corresponding to 
\begin{equation}
\lambda_j>1+\frac{1}{2\sqrt{n}}.
\end{equation}
The probability of finding an eigenvalue in $(1+\frac{1}{2\sqrt{n}}, +\infty)$ is tiny.  As we shall show in Corollary 2
\begin{equation}
P\left (\#\left \{\lambda_j>1+\frac{1}{2\sqrt{n}}\right \}>0\right )\leq c_1\cdot\exp (-c_2\cdot n^{\frac{1}{6}}).
\end{equation}
where $c_1,c_2$ are some positive constants.  The estimate (1.26) implies that for the purpose of calculating the limiting $k$-point correlation functions,
it is enough to consider $S_{n,k}(t_1,\ldots ,t_k)$, rather than the whole sum (1.22).  The following two results will be proved in \S 5.
\medskip

\noindent{\bf Lemma 9}.
{\it
The mathematical expectation of $S_{n,k}(t_1,\ldots ,t_k)$ converges to the limit (1.23) as $n\rightarrow\infty$:
$E\ S_{n,k}(t_1,\ldots t_k)
\underset{n\rightarrow\infty}{\longrightarrow} 
\int^\infty_{-\infty}\cdots \int^\infty_{-\infty}\exp (t_1\cdot y_1+\cdots +t_ky_k)\cdot R_{\beta ,k}(y_1,\ldots ,y_k) dy_1\cdots dy_k$ for any $t_1,t_2,\dots t_k>0$, where $R_{\beta ,k}, k=1,2,\dots ,$ is the limiting $k$-point correlation function of the G.U.E. ($\beta =2$) and G.O.E. ($\beta =1$).}
\medskip

\noindent{\bf Theorem A}
{\it 
Let $A_n$ be a Wigner random hermitian ((1.1)--(1.3)) or real symmetric ((1.1')--(1.3')) matrix.  Then rescaled correlation functions at the edge weakly converge to the universal limits (G.U.E. or G.O.E., correspondingly) as $n\rightarrow\infty$.}
\medskip

In [22], [23] a special combinatorial technique has been developed to treat statistics (1.24).  The main idea is to study traces of high powers of $A_n$.  Let us consider
$$\Trace\ A_n^{2[t\cdot n^{\frac{2}{3}}]}=\sum^n_{j=1}\lambda_j^{2s_n},$$
where $s_n=[t\cdot n^{\frac{2}{3}}]$ and $t>0$.  Define rescaling at the edge of the spectrum by $\lambda_j=1+\frac{\theta_j}{2n^{\frac{2}{3}}}$ for positive eigenvalues $\lambda_j>0$, and $\lambda_j=-1-\frac{\tau_j}{2n^{\frac{2}{3}}}$ for $\lambda_j\leq 0$.  Then
$$\sum^n_{j=1}\lambda_j^{2[t\cdot n^{\frac{2}{3}}]}=
\sum_j\,'\left (1+\frac{\theta_j}{2n^{\frac{2}{3}}}\right )^{2[t\cdot n^{\frac{2}{3}}]}+
\sum_j\,''\left (-1-\frac{\tau_j}{2n^{\frac{2}{3}}}\right )^{2[t\cdot n^{\frac{2}{3}}]},
$$
where summation in $\sum'_j$ is over $\lambda_j>0$ and in $\sum_j''$ is over $\lambda_j\leq 0$.  To proceed we break both $\sum'$ and $\sum''$ into three subsums: $\sum'=I'_1+I'_2+I'_3$, where 
\begin{align*}
I'_1&=~\sum_{j: 0<\lambda_j<1-\frac{1}{2n^\frac{1}{2}}}
(1+\frac{\theta_j}{2n^{\frac{2}{3}}})^{2[t\cdot n^{\frac{2}{3}}]},\\ 
I'_2&=\sum_{j:1-\frac{1}{2n^\frac{1}{2}}\leq\lambda_j\leq 1+\frac{1}{2n^{\frac{1}{2}}}}
(1+\frac{\theta_j}{2n^{\frac{2}{3}}})^{2[t\cdot n^{\frac{2}{3}}]},\\
I'_3&=\sum_{j:\lambda_j>1+\frac{1}{2n^{\frac{1}{2}}}}
(1+\frac{\theta_j}{2n^{\frac{2}{3}}})^{2[t\cdot n^{\frac{2}{3}}]}
\end{align*}
and similarly $\sum''=I''_1\ +\ I''_2\ +\ I''_3$, where
\begin{align*}
I''_1&=\sum_{j:-1+\frac{1}{2n^{\frac{1}{2}}}<\lambda_j\leq 0}
\left (-1-\frac{\tau_j}{2n^{\frac{2}{3}}}\right )^{2[t\cdot
n^{\frac{2}{3}}]},\\
I''_2&=\sum_{j:-1-\frac{1}{2n^{\frac{1}{2}}}\leq\lambda_j\leq -1
+\frac{1}{2n^{\frac{1}{2}}}}
\left (-1-\frac{\tau_j}{2n^{\frac{2}{3}}}\right )^{2[t\cdot 
n^{\frac{2}{3}}]},\\
I''_3&=\sum_{j:\lambda_j<-1-\frac{1}{2n^{\frac{1}{2}}}}
\left (-1-\frac{\tau_j}{2n^{\frac{2}{3}}}\right )^{2[t\cdot n^{\frac{2}{3}}]}.
\end{align*}

By definition, the sum in $I'_2,\ I''_2$ is over $\vert\theta_j\vert\leq n^{\frac{1}{6}},\ \vert\tau_j\vert\leq n^{\frac{1}{6}}$, while in $I'_1,I''_1 (I'_3,I''_3)$ the sum is over $-2n^{\frac{2}{3}}<\theta_j<-n^{\frac{1}{6}},\ 
-2n^{\frac{2}{3}}\leq\tau_j<-n^{\frac{1}{6}}\ (\theta_j>n^{\frac{1}{6}},\ \tau_j>n^{\frac{1}{6}})$.  One can see that $I'_1,\ I''_2$ are going to zero as $n\rightarrow\infty$:
\begin{equation}
0\leq I'_1,\ I''_1\leq n\cdot \left (1-\frac{1}{2n^{\frac{1}{2}}}\right )^{2[tn^{\frac{2}{3}}]}\leq
n\cdot\exp \left ( -\frac{t}{2}n^{\frac{1}{6}}\right ).
\end{equation}
If we look at $I'_2,\ I''_2$ then uniformly in $t$ from compact subsets of $(0, +\infty)$ we have
\begin{align}
\begin{split}
&I'_2=\sum_{\vert\theta_j\vert\leq n^{\frac{1}{6}}}
e^{t\cdot\theta_j}\cdot\left (1+\underbar{0}(n^{-\frac{1}{3}})\right )\\
&I''_2=\sum_{\vert\tau_j\vert\leq n^{\frac{1}{6}}}
e^{t\cdot\tau_j}\cdot\left (1+\underbar{0}(n^{-\frac{1}{3}})\right ).
\end{split}
\end{align}
Finally, let us restrict our attention to $I'_3,\ I''_3$.  Such subsums correspond to the large deviations of maximal (minimal) eigenvalues and with probability greater than $1-c_1\cdot\exp (-c_2\cdot n^{\frac{1}{6}})$ will be shown to contain no terms at all.  The above arguments combined with some technical considerations in \S 2 and 4 will imply
\begin{equation}
\Trace\ A^{2[tn^{\frac{2}{3}}]}-\left ( \sum_{\theta_j<n^{\frac{1}{6}}}e^{t\cdot\theta_j}+
\sum_{\tau_j<n^{\frac{1}{6}}}e^{t\cdot\tau_j}\right )
\underset{n\rightarrow\infty}{\longrightarrow} 0\ \text{(a.e.)}.
\end{equation}
We shall also be able to establish that all the moments of the l.h.s.~of (1.29) converge to zero.  Considering Trace $A^{2[tn^{\frac{2}{3}}]+1}$, similar arguments will provide
\begin{equation}
\Trace\ A^{2[tn^{\frac{2}{3}}]+1}-\left ( \sum_{\theta_j<n^{\frac{1}{6}}}e^{t\cdot\theta_j}
-\sum_{\tau_j<n^{\frac{1}{6}}}e^{t\cdot\tau_j}\right )
\underset{n\rightarrow\infty}{\longrightarrow}\ 0,
\end{equation}
where the convergence is again with probability 1 as well as in $L^p,\ p\geq 1$.  It follows from (1.29), (1.30) that 
$$\sum_{\theta_j<n^{\frac{1}{6}}}e^{t\cdot\theta_j}-\frac{1}{2}\cdot (\Trace\ A^{2[tn^{\frac{2}{3}}]}+
\Trace\ A^{2[tn^{\frac{2}{3}}]+1})\underset{n\rightarrow\infty}
{\longrightarrow}\ 0
\ \ \text{(a.e.)}.$$
Linear statistics $\sum_{\theta_j<n^{\frac{1}{6}}}e^{t\cdot\theta_j},\ \sum_{\tau_j<n^{\frac{1}{6}}}e^{t\cdot\tau_j}$ are identically distributed (this follows from the symmetry $A\rightarrow -A$ of the model).  Once we prove in \S 3, 4 that traces of consecutive high powers are asymptotically independent, this will imply asymptotical independence of the linear statistics.  An easy way to explain this is to say that distributions of eigenvalues in far apart regions ($\{\lambda :\vert\lambda -1\vert <\frac{1}{2n^{\frac{1}{2}}}\}$
and $\{\lambda :\vert\lambda +1\vert< \frac{1}{2n^{\frac{1}{2}}}\}$ in our case) are independent in the limit $n\rightarrow \infty$.  Of course the actual proof requires some work.

In [22], [23] we studied the traces of high powers $A^{p_n}_n,\ p_n=2[t_n\cdot n^{\frac{2}{3}}],\ 2[t_n\cdot n^{\frac{2}{3}}]+1$, under the assumption 
$t_n\underset{n\rightarrow\infty}{\longrightarrow}0$.
In particular, we established

\noindent{\bf Theorem} (Sinai, Soshnikov)
{\it 
Let $p_n\rightarrow +\infty$ be such that $p_n/n^{\frac{2}{3}}\rightarrow 0$ as $n\rightarrow +\infty$.  Then
$$E(\Trace\ A^{p_n}_n)=\begin{cases}\sqrt{\frac{8}{\pi}}\cdot \frac{n}{p_n^{\frac{3}{2}}}\cdot (1+\bar o(1)) & \text{if $p_n$ even},\\
0 & \text{if $p_n$ odd},\end{cases}$$
and the moments of the centralized trace \Trace $A^{p_n}_n-E\ \Trace A^{p_n}_n$ converge to the moments of $N(0,\ \frac{1}{\pi})$.}
\medskip

Similar results have been established for the joint $k$-dimensional distribution of Trace $A^{p_{n_1}},\ldots ,\Trace\ A^{p_{n_k}}$ provided that $p_{n_1},\ldots , p_{n_k}$ are of the same order.

The theorem itself says nothing about the moments of Trace $A^{2[t\cdot n^{\frac{2}{3}}]}$ for fixed $t>0$.  However as an easy corollary we have
\medskip

\noindent{\bf Theorem 1}
{\it
For any $\epsilon >0, K>0$ there exists some $\delta (\epsilon ,K)>0$ such that if $0<t<\delta ,\ 1\leq k\leq K$, then the $k$th moment of 
Trace $A^{2[t\cdot n^{\frac{2}{3}}]}$ 
stays bounded as $n\rightarrow\infty$ and}
\begin{equation}
\begin{split}
(1-\epsilon )\cdot\pi^{-\frac{k}{2}}\cdot t^{-\frac{3k}{2}}\liminf_{n\rightarrow\infty}\ E(\Trace\ 
A^{2[t\cdot n^{\frac{2}{3}}]})^k
&\leq\limsup_{n\rightarrow\infty}\ E(\Trace\ 
A^{2[t\cdot n^{\frac{2}{3}}]})^k\\
&\leq (1+\epsilon )\cdot\pi^{-\frac{k}{2}}\cdot t^{-\frac{3k}{2}}.
\end{split}
\end{equation}
\medskip

\noindent{\bf Proof}.
Suppose (1.31) is false.  Then there exists a sequence $t_m\searrow 0$ such that $
\limsup_{n\rightarrow\infty}\ E(\Trace\ 
A^{2[t\cdot n^{\frac{2}{3}}]})^k>
(1+\epsilon )\cdot\pi^{-\frac{k}{2}}\cdot t_m^{-\frac{3k}{2}}$ (we shall concentrate here only on the $\limsup$ part, since 
the $\liminf$ part is similar).  Then for each $m$ there exists sufficiently large $n_m$ such that
\begin{equation}
E(\Trace\ 
A_{n_m}^{2[t_m\cdot n_m^{\frac{2}{3}}]})^k
>(1+\frac{\epsilon}{2})\cdot
\pi^{-\frac{k}{2}}\cdot t_m^{-\frac{3k}{2}}.
\end{equation}
We can choose $n_m$ so that $t_m\cdot n_m^{\frac{2}{3}}\rightarrow +\infty$.
Now taking $p_{n_m}=2[t_m\cdot n_m^{\frac{2}{3}}]$ one concludes that (1.32) contradicts the convergence of $E(\Trace\ A_{n_m}^{2[t_m\cdot n_m^{\frac{2}{3}}]}-\frac{1}{\sqrt\pi}\cdot t_m^{-\frac{3}{2}})^k$ to the $k$th moment of $N(0,\frac{1}{\pi})$.\qed

The analogue of the theorem for the joint distribution of 
Trace $A^{2[t_n\cdot n^{\frac{2}{3}}]}$, Trace $A^{2[t_n\cdot n^{\frac{2}{3}}]+1}$ says that (Trace $A^{2[t_n\cdot n^{\frac{2}{3}}]}-
E\ \Trace\ A^{2[t_n\cdot n^{\frac{2}{3}}]}$, Trace $A^{2[t_n\cdot n^{\frac{2}{3}}]+1})\break\underrightarrow{w}$ $N(0,\ \frac{1}{\pi}\cdot Id)$ (see [22], [23]) provided $t_n\rightarrow 0$ and $t_n\cdot n^{\frac{2}{3}}\rightarrow +\infty$.

Since
\begin{align}
\begin{split}
&\sum_{\theta_j<n^{\frac{1}{6}}}e^{t\theta_j}=\frac{1}{2}(\Trace\ A^{2[t_n\cdot n^{\frac{2}{3}}]}\\
&\qquad +\Trace\ A^{2[t_n\cdot n^{\frac{2}{3}}]+1})\cdot (1+\underline{0}(n^{-\frac{1}{3}}))+\bar o(1)
\end{split}
\end{align}

\begin{align}
\begin{split}
&\sum_{\tau_j<n^{\frac{1}{6}}}e^{t\tau_j}=\frac{1}{2}(\Trace\ A^{2[t_n\cdot n^{\frac{2}{3}}]}\\
&\qquad -\Trace\ A^{2[t_n\cdot n^{\frac{2}{3}}]+1})\cdot (1+\underline{0}(n^{-\frac{1}{3}}))+\bar o(1)\ \ \text{(a.e.)}
\end{split}
\end{align}
One may hope to establish results similar to Theorem 1 for the linear statistics.  This is indeed the case:
\medskip

\noindent{\bf Corollary 1}
{\it
For any $\epsilon >0, K>0$ there exists $\delta (\epsilon ,K)>0$ such that if $0<t<\delta$, $1\leq k\leq K$ then the $k$th moments of $\sum_{\theta_j<n^{\frac{1}{6}}}e^{t\cdot\theta_j},\ \sum_{\tau_j<n^{\frac{1}{6}}}e^{t\cdot\tau_j}$ stay bounded as $n\rightarrow\infty$ and}
\begin{align}
\begin{split}
&(1-\epsilon )\cdot\left (\frac{1}{2\sqrt\pi}\right )^k\cdot t^{-\frac{3k}{2}}\leq\liminf_{n\rightarrow\infty}\ E\left (\sum_{\theta_j<n^{\frac{1}{6}}}e^{t\theta_j}\right )^k\\
&\qquad \leq\limsup_{n\rightarrow\infty}\ E\left (\sum_{\theta_j<n^{\frac{1}{6}}}e^{t\theta_j}\right )^k\leq
(1+\epsilon )\cdot\left (\frac{1}{2\sqrt\pi}\right )^k\cdot t^{-\frac{3k}{2}},
\end{split}
\end{align}

\begin{align}
\begin{split}
&(1-\epsilon )\cdot\left (\frac{1}{2\sqrt\pi}\right )^k\cdot t^{-\frac{3k}{2}}\leq\liminf_{n\rightarrow\infty}\ E\left (\sum_{\tau_j<n^{\frac{1}{6}}}e^{t\tau_j}\right )^k\\
&\qquad\leq\limsup_{n\rightarrow\infty}\ E\left (\sum_{\tau_j<n^{\frac{1}{6}}}e^{t\tau_j}\right )^k\leq
(1+\epsilon )\cdot\left (\frac{1}{2\sqrt\pi}\right )^k\cdot t^{-\frac{3k}{2}}.
\end{split}
\end{align}
\medskip

\noindent{\bf Proof}
It follows from the arguments around the formulas (1.27), (1.28) that
\begin{align*}
\begin{split}
&\left\vert\sum_{\theta_j<n^{\tfrac{1}{6}}}e^{t\cdot\theta_j}-
\tfrac{1}{2}\cdot\Trace\ A^{2[t\cdot n^{\frac{2}{3}}]}-
\frac{1}{2}\cdot \Trace\ A^{2[t\cdot n^{\frac{2}{3}}]+1}\right\vert\leq\\
& \qquad r_1+r_2+r_3,
\end{split}
\end{align*}
where
\begin{align*}
\begin{split}
r_1&=\const\cdot n^{-\frac{1}{3}}\cdot (\Trace\ A^{2[t\cdot n^{\frac{2}{3}}]}
+
\vert \Trace\ A^{2[t\cdot n^{\frac{2}{3}}]+1}\vert ),\\
r_2&=\sum_{\lambda_j>1+\frac{1}{2n^{\frac{1}{2}}}}\lambda_j^{{2[t\cdot n^{\frac{2}{3}}]+1}},\\
r_3&=\sum_{\lambda_j<-1-\frac{1}{2n^{\frac{1}{2}}}}-\lambda_j^{{2[t\cdot n^{\frac{2}{3}}]+1}}.
\end{split}
\end{align*}
Then it is enough to show that for any given $K>0$, the first $K$ moments of $r_1,\ r_2,\ r_3$ vanish as $n\rightarrow\infty$, provided $t$ is small enough.  The statement for $r_1$ immediately follows from Theorem 1.  To consider $r_2$, we write
$r_2\cdot (1+\frac{1}{2n^{\frac{1}{2}}})^{2[t\cdot
n^{\frac{2}{3}}]-1}\leq \Trace\ A^{2[2t\cdot n^{\frac{2}{3}}]}$.
Therefore 
$$r_2\leq \left (1+\frac{1}{2n^{\frac{1}{2}}}\right )^{1-2[t\cdot
n^{\frac{2}{3}}]}\cdot \Trace\ A^{2[2t\cdot n^{\frac{2}{3}}]}\leq e^{-\frac{t}{2}n^{\frac{1}{6}}}\cdot \Trace\ A^{2[2t\cdot n^{\frac{2}{3}}]}$$
for sufficiently large $n$.  Now taking $2t<\delta (\epsilon ,K)$ and
applying Theorem 1, one can show that the first $K$ moments of $r_2$ vanish as $n\rightarrow\infty$.  The case of $r_3$ can be done in a similar fashion.\qed
\medskip

As another corollary of Theorem 1, we prove the estimate (1.26):
\medskip

\noindent{\bf Corollary 2}
{\it
$$P(\#\{\lambda_j>1+\frac{1}{2n^{\frac{1}{2}}}\}>0)\leq c_1\cdot\exp (-c_2\cdot n^{\frac{1}{6}})$$
where $c_1,\ c_2$ are some positive constants.}

\noindent{\bf Proof}

$$P\left (\#\{\lambda_j>1+\frac{1}{2n^{\frac{1}{2}}}\}>0\right )\leq \left (E\sum_{\lambda_j>1+\frac{1}{2n^{\frac{1}{2}}}}
\lambda_j^{2[\frac{\delta}{2}n^{\frac{2}{3}}]}\right )\cdot
(1+\frac{1}{2n^{\frac{1}{2}}})^{-2[\frac{\delta}{2}n^{\frac{2}{3}}]}
$$
with $\delta =\delta (1,1)$ from Theorem 1.  Then the last inequality implies
$$P\left (\#\{\lambda_j>1+\frac{1}{2n^{\frac{1}{2}}}\}>0\right )
\leq 2\cdot\pi^{-\frac{1}{2}}\cdot \left (\frac{\delta}{2}\right )^{-\frac{3}{2}}\cdot e^{-\frac{\delta}{2}\cdot n^{\frac{1}{6}}}.$$

\noindent{\it Remark 5}

It will follow from our results in the \S 2--5 that 
$$E\left (\sum_{\theta_j<n^{\frac{1}{6}}}e^{t\theta_j}\right )^k,\ 
E\left (\sum_{\tau_j<n^{\frac{1}{6}}}e^{t\cdot\tau_j}\right )^k$$
have limits for any $t>0$ and $k\in\bbZ^1_+$.

The rest of the paper is organized as follows.  In \S 2 we formulate Theorems 2 and 3 together with some corollaries and revisit a combinatorial problem associated with calculation of moments of Trace $A^{p_n}$.  This problem deals with the number of closed paths of length $p_n$ on a complete nonoriented graph with $n$ vertices, where the paths are such that each edge appears an even number of times.  In [22], [23] we focused on the properties of typical paths when $p_n/n^{\frac{2}{3}}\rightarrow 0$.  The
case when $p_n$ is proportional to $n^{\frac{2}{3}}$, which is of at most
importance to the statistics at the edge, will be treated in this paper.  For a warm-up we shall consider a simpler problem in \S 3:  the dependence
of the behavior of typical closed paths on $p_n$ when there is no additional condition that every edge appears in the path an even number of times.
Theorems 2, 3 are proven in \S 4.  In \S 5 we deduce Theorem A from Theorems 2, 3 and prove the main result.

It is a great pleasure to thank Ya. Sinai for his inspiration and interest in this work.  The author also would like to thank P. Forrester for the warm hospitality at the University of Melbourne in July 1997 and valuable discussions.

The research was partially supported by the National Science Foundation through Grant No. DMS-9304580.

\section{Traces of High Powers of Wigner Matrices}

We start with a calculation of the mathematical expectation of Trace 
$A^{p_n}_n$, where $p_n=2s_n$ or $2s_n+1$; $s_n=[t\cdot n^{\frac{2}{3}}]$.
Clearly,
\begin{equation}
E\ \Trace\ A^{P_n}_n=\sum_{\p}E\ a_{i_0,i_1}a_{i_1,i_2}\dots a_{i_{p_n-1}}i_0.
\end{equation}
The sum in (2.1) is taken over all closed paths $\p =\{i_0,i_1,\ldots ,i_{p_n-1},i_0\}$, with a distinguished origin, in the set $\{1,2,\dots n\}$.  We consider the set of vertices $\{1,2,\dots n\}$ as a nonoriented graph in which any two vertices are joined by an unordered edge.  Since the distributions of the random variables $a_{ij}$ are symmetric, we conclude that the only paths giving nonzero contribution to (2.1) are those for which the number of occurrences of each edge is even.  Indeed, due to the independence of $\{ a_{ij}\}_{i\leq j}$, the mathematical expectation of the product factorizes as a product of mathematical expectations of random variables corresponding to different edges of the path.  Therefore if some edge appears in $\p$ odd number of times at least one factor in the product will be zero.  In particular, if the length of $\p$ is odd ($P_n=2[t\cdot n^{\frac{2}{3}}]+1)$, then
$$E\ \Trace A^{2[t\cdot n^{\frac{2}{3}}]+1}=0.$$ 
For the even powers of $A$, we established in Theorem 1 that $E\ \Trace ^{2[t\cdot n^{\frac{2}{3}}]}$ is uniformly bounded in $n$ for sufficiently small $t$.  We shall generalize this result in the next theorem:
\medskip

\noindent{\bf Theorem 2}
{\it 
Let $A_n$ be either a hermitian ((1.1)--(1.3)) or real symmetric 
((1.1')--(1.3')) Wigner random matrix.  Then the following is true:
\begin{enumerate}
\item[a)] There are some constants $\gamma_1,\ \gamma_2>0$ such that for any
$t>0$
\begin{equation}
E\ \Trace A^{2[t\cdot n^{\frac{2}{3}}]}\leq\frac{\gamma_1}{t^{\frac{3}{2}}}
e^{\gamma_2t^3}
\end{equation}
for sufficiently large $n$ (depending on $t$).

\item[b)] A subsum of (2.1) that corresponds to the paths, where either at least one edge appears more than twice or there are loops (edges $\{j,j\},j=1,\dots n$), goes to zero as $n\rightarrow +\infty$.
\end{enumerate}
}

We will prove Theorem 2 in the \S 4.  A remarkable corollary of it consists of the fact that the limit of $E\ \Trace A^{2[t\cdot n^{\frac{2}{3}}]}$ exists for an arbitrary Wigner matrix and is the same as in the special case of G.U.E. (G.O.E.).

We start with
\medskip

\noindent{\bf Lemma 1}.
{\it
For the Gaussian Unitary Ensemble $\lim_{n\rightarrow\infty}E\ A^{2[t\cdot n^{\frac{2}{3}}]}$ exists for all $t>0$ and is equal to $2\cdot\int^\infty_{-\infty}e^{t\theta}R_{2,1}(\theta )\, d\theta$.}
\medskip

\noindent{\bf Lemma 2}.
{\it
For the Gaussian Orthogonal Ensemble, $\lim_{n\rightarrow\infty}E\ A^{2[t\cdot n^{\frac{2}{3}}]}$ exists for all $t>0$ and is equal to $2\cdot\int^\infty_{-\infty}e^{t\theta}R_{1,1}(\theta )\, d\theta$.}
\medskip

\noindent{\it Remark 6}

The limits for the hermitian and real symmetric case are different, which is not a surprise since local statistics at the edge (e.g. correlation functions) are different for G.U.E. and G.O.E.

We shall prove here Lemma 1 only.  The proof for Lemma 2 is quite similar.

\noindent{\bf Proof of Lemma 1}

The main ingredient of the proof is the claim that mathematical expectations of linear statistics, $\sum_j e^{t\theta_j},\ \sum_{j: \theta_j<n^{\frac{1}{6}}}  e^{t\theta_j},\ \sum_je^{t\cdot\tau_j},\ \sum_{j:\tau_j<n^{\frac{1}{6}}}
e^{t\cdot\tau_j}$ have the same limit as $n\rightarrow\infty$.  To calculate
mathematical expectations of linear statistics, we need to know the exact formula
for the spectral density (one-point correlation function), which in the case
of G.U.E. is equal to 
\begin{equation}
\rho_{n,2,1}(x)={\sqrt {2n}}\cdot\sum^{n-1}_{l=0}\psi^2_\ell (\sqrt{2nx)},
\end{equation}
(see [17]) where
\begin{equation}
\psi_\ell (x)=\frac{(-1)^\ell}{\pi^{\frac{1}{4}}\cdot (2^\ell\cdot\ell 
!)^{\frac{1}{2}}}\exp\left (\frac{x^2}{2}\right )\frac{d^\ell}{dx^\ell}
(\exp (-x^2))
\end{equation}
are known as Weber-Hermite functions or normalized eigenfunctions of harmonic
oscillator:
\begin{equation}
\left (-\frac{1}{2}\frac{d^2}{dx^2}+\frac{x^2}{2}\right )\psi_\ell =
\left (\ell +\frac{1}{2}\right )\psi_\ell ,\ \ \ \ell=0,1,\dots .
\end{equation}
For use later, we also write here formulas for $k$-point correlation
functions:
\begin{equation}
\rho_{n,2,k}(x_1,\dots ,x_k)=\det (K_n(x_i,x_j))^n_{i,j=1},
\end{equation}
where
\begin{equation}
K_n(x,y)=\sqrt{2n}\cdot\sum^{n-1}_{\ell =0}\psi_\ell (\sqrt{2nx})\cdot
\psi_\ell (\sqrt{2ny}).
\end{equation}
It follows from the asymptotics of Hermite polynomials ([49]) that for
$x=1+\frac{\theta}{2n^{\frac{2}{3}}}$ we have
\begin{equation}
\lim_{n\rightarrow\infty} n^{\frac{1}{12}}\psi_n(\sqrt{2nx})=2^{\frac{1}{4}}
\a_i(\theta )
\end{equation}
uniformly in $\theta$ bounded from below, and for large positive $\theta$ uniformly in $n$
\begin{equation}
n^{\frac{1}{12}}\psi_n(\sqrt{2nx})=\bar o(e^{-\theta}).
\end{equation}
Equations (2.8), (2.9) imply that for $\theta$ bounded from below we have uniform convergence
\begin{equation}
\lim_{n\rightarrow\infty}\frac{1}{2n^{\frac{2}{3}}}\rho_{n,2,1}\left (1+\frac{\theta}{2n^{\frac{2}{3}}}\right )=\int^\infty_0\a_i(\theta +t)^2dt
\end{equation}
and similarly for higher correlation functions
\begin{equation}
\lim_{n\rightarrow\infty}\left (\frac{1}{2\cdot n^{\frac{2}{3}}}\right )\cdot\rho_{n,2,k}\left (
1+\frac{\theta_1}{2n^{\frac{2}{3}}},\ldots ,1+\frac{\theta_k}{2n^{\frac{2}{3}}}\right )
=\det (K(\theta_j,\ \theta_j))^k_{i,j=1}
\end{equation}
uniformly in $\theta_1,\dots ,\theta_k$ bounded from below, where
\begin{align}
\begin{split}
K(x,\ y)&=\int^\infty_0\a_i(x+t)\a_i(y+t)\, dt\\
&=\frac{\a_i(x)\cdot\a_i'(y)-\a_i'(x)\cdot\a_i(y)}
{x-y}.
\end{split}
\end{align}
We denote the limit in (2.11) by $R_{2,k}(\theta_1,\ldots \theta_k)$.

As an immediate consequence of (2.9), we have
\begin{equation}
E\sum_{\theta_j\geq n^{\frac{1}{6}}}e^{t\cdot\theta_j}=
\int^{+\infty}_{n^{\frac{1}{6}}}e^{t\cdot\theta}\cdot R_{n,2,1}(\theta )\, d\theta 
\underset{n\rightarrow\infty}{\longrightarrow} 0.
\end{equation}
For arbitrary positive $T$, formulas (2.8), (2.9) imply
\begin{equation}
E=\sum_{\theta_j\geq -T}e^{t\cdot\theta_j}=\int^{+\infty}_{-T}e^{t\theta}\cdot R_{n,2,1}
(\theta )\, d\theta \underset{n\rightarrow\infty}{\longrightarrow} \int^{+\infty}_{-T}e^{t\cdot
\theta}R_{2,1}(\theta )\, d\theta .
\end{equation}
To prove the convergence
\begin{equation}
E\sum_je^{t\cdot\theta_j}\underset{n\rightarrow\infty}{\longrightarrow} \int^\infty_{-\infty}
e^{t\theta}\cdot R_{2,1}(\theta )\, d\theta .
\end{equation}
we have to justify taking the limit in (2.14).

One way to do this is by using Plancherel-Rotach asymptotics of Hermite functions near the turning point for large negative $\theta$.  Or one can do it as follows:

Let $\delta =\delta (1,1)$ be as from Corollary 1 (\S 1).  Take $\varkappa =\min (\frac{t}{2},\ 
\frac{\delta}{2})$.  Then
\begin{align}
\begin{split}
E\sum_{\theta_j<-T}e^{t\cdot\theta_j}&\leq E\sum_{\theta_j<-T}e^{\varkappa
\cdot\theta_j}\cdot e^{-\varkappa\cdot T}\\ 
&\leq E\sum_{\theta_j<n^{\frac{1}{6}}}e^{\varkappa\cdot\theta_j}\cdot
e^{-\varkappa\cdot T}\leq
\frac{1}{\sqrt\pi}\varkappa^{-\frac{3}{2}}\cdot e^{-\varkappa\cdot T}
\underset{T\rightarrow\infty}{\longrightarrow} 0
\end{split}
\end{align}
uniformly in $n$.

Combining (2.13), (2.14) and (2.16), we obtain
$$\lim_{n\rightarrow\infty}E\sum_je^{t\theta_j}=\lim_{n\rightarrow\infty}E
\sum_{\theta_j<n^{\frac{1}{6}}}e^{t\theta_j}=
\int^\infty_{-\infty}e^{t\theta}R_{2,1}(\theta )\, d\theta .$$
Because of the symmetry $A\rightarrow -A$ of the model, we also conclude that the last equation
holds for $E\sum_je^{t\cdot\tau_j}$ and $E\sum_{\tau_j<n^{\frac{1}{6}}}e^{t\cdot\tau_j}$.

To conclude the proof of the Lemma 1 we note that formulas (1.27), (1.28) and discussions 
around them claim
\begin{align*}
\begin{split}
&\left \vert E\ \Trace\ A^{2[t\cdot n^{\frac{2}{3}}]}-E\sum_{\theta_j<n^{\frac{1}{6}}}
e^{t\cdot\theta_j}-
E\sum_{\tau_j<n^{\frac{1}{6}}}e^{t\cdot\tau_j}\right \vert\leq\\
&\qquad\frac{\const}{n^{\frac{1}{3}}}\cdot
\left (E\sum_{\theta_j<n^{\frac{1}{6}}}e^{t\cdot\theta_j}+
E\sum_{\tau_j<n^{\frac{1}{6}}}e^{t\cdot\tau_j}\right )+E\sum_{\theta_j\geq n^{\frac{1}{6}}}
e^{t\cdot\theta_j}+\\
&\qquad +E\sum_{\tau_j> n^{\frac{1}{6}}}e^{t\cdot\tau_j}+2n\cdot\exp \left (-\frac{t}{2}n^{\frac{1}{6}}\right ).
\end{split}
\end{align*}
To finish the proof, we note that the r.h.s. in the last inequality goes to zero.\qed

A similar strategy works for the G.O.E. too and allows us to establish Lemma 2.  For the formulas for $k$-point correlation functions in the G.O.E., the reader is referred to [17], [27].

After we formulated Theorem 2 and proved Lemmas 1 and 2, the picture starts to emerge.
Assuming we can prove Theorem 2 (we shall do it in \S 4), we arrive at the following corollaries.
\medskip

\noindent{\bf Corollary 3}
{\it
For an arbitrary Wigner hermitian matrix (1.1)-(1.3) the limit of $E\ \Trace\ 
A_n^{2[t\cdot n^{\frac{2}{3}}]}$ exists for all $t>0$ and coincides with the
limit for the G.U.E. from Lemma 1.}
\medskip

\noindent{\bf Corollary 4}
{\it
For an arbitrary Wigner real symmetric matrix (1.1')-(1.3') the limit of $E\ \Trace\ 
A_n^{2[t\cdot n^{\frac{2}{3}}]}$ exists for all $t>0$ and coincides with the limit for G.O.E. from Lemma 2.}
\medskip

The proof of the corollaries is elementary.  Subsums of (2.1) over the paths such that each edge appears in the path twice or does not appear at all and there are no loops (i.e., edges of the form $(i, i)$) depend only on the second moments of $\{a_{ij}\}_{1\leq i<j\leq n}$ that are the same for all Wigner matrices (within its symmetry class).  Since the rest of the sum (2.1) goes to zero according to part b) of Theorem 2, Corollaries 1, 2 are proven.\qed

To study the higher moments of Trace $A^{p_n},\ p_m=2[t\cdot n^{\frac{2}{3}}],\ 2[t\cdot n^{\frac{2}{3}}]+1$, we write similarly to (2.1)
\begin{equation}
E\prod^k_{j=1}\ \Trace\ A^{P^{(j)}_n}_n=\sum_{\p_1,\dots \p_k}E\prod^k_{j=1}
\prod^{p^{(j)}_n-1}_{l=0}a_{i_\ell^{(j)}\ i^{(j)}_{\ell +1}},
\end{equation}
where the sum is taken over all closed paths $\p_j=\{i^{(j)}_0,\ i^{(j)}_1,
\ldots 
,i^{(j)}_{p^{(j)}_n-1},\ i^{(j)}_0\}$, $j=1,\dots k$, in the set of $n$ vertices.  Since each path $\p_j$ is closed its distinguished origin $i^{(j)}_0$ coincides with the end vertex
$i^{(j)}_{p^{(j)}_n}$.  The next theorem will be proven in \S 4.  It establishes an analogue of Theorem 2 for higher moments.
\medskip

\noindent{\bf Theorem 3}
{\it 
Let $A_n$ be either a hermitian ((1.1)--(1.3)) or real symmetric 
((1.1')--(1.3')) Wigner random matrix.  Then there are some constants $\gamma_1,\ \gamma_2>0$ such that for any $t_1,\ t_2,\dots t_k>0$ and
$$p^{(1)}_n\in\{2[t_1\cdot n^{\frac{2}{3}}],\ 2[t_1\cdot n^{\frac{2}{3}}]+1\},
\ldots ,p^{(k)}_n\in \{2[t_k\cdot n^{\frac{2}{3}}],\ 2[t_k\cdot n^{\frac{2}{3}}]
+1\},$$
the following holds:
\begin{enumerate}
\item[a)]
\begin{equation}
E\prod^k_{i=1}\ \Trace\ A^{p^{(i)}_n}_n\leq\frac{\gamma^k_1}{\prod^k_{i=1}
t_i^{3k/2}}\cdot\exp (\gamma_2\cdot\sum^k_{i=1}t^3_i)
\end{equation}
for sufficiently large $n$ (depending on $t_1, \ldots, t_k$).
\item[b)] A subsum of (2.18) over $k$-tuples of paths $(\p_1,\ldots ,\p_k)$ for which at least one nonoriented edge appears more than twice in their union
or at least one path has a loop, vanishes in the limit $n\rightarrow\infty$.
\end{enumerate}
}
Let us consider first the Gaussian Ensembles.
\medskip

\noindent{\bf Lemma 3}  
{\it
For the Gaussian Unitary Ensemble, the mathematical expectation at the 
l.h.s.~of (2.18) has a limit as $n\rightarrow\infty$.}

Similarly, for the G.O.E.:
\medskip

\noindent{\bf Lemma 4}
{\it
For the Gaussian Orthogonal Ensemble the mathematical expectation at the l.h.s. of (2.18) has a limit as $n\rightarrow \infty$.}

The proof of Lemmas 3 and 4 essentially follows from the existence of limiting $k$-point correlation functions in G.U.E. and G.O.E.  All we have to do is to consider linear statistics $\sum_j\exp (t_i\cdot\theta_j),\ \sum_{j:\theta_j<n^{\frac{1}{6}}}\exp (t_i\cdot\theta_j),\ 
\sum_{j:\tau_j<n^{\frac{1}{6}}}\exp (t_i\cdot\tau_j),\ 
\sum_{j:\tau_j<n^{\frac{1}{6}}}\exp (t_i\cdot\tau_j),
i=1,\dots k$, and note that
\begin{enumerate}
\item[(i)] their moments have limits as $n\rightarrow\infty$;

\item[(ii)] $\sum_j\exp (t_i\cdot\theta_j),\ \sum_j\exp (t_m\cdot\tau_j)$ are asymptotically independent, and

\item[(iii)] the moments of $\sum_{j:\theta_j\geq n^{\frac{1}{6}}}\exp (t_i\cdot\theta_j),\ \sum_{j:\tau_j\geq n^{\frac{1}{6}}}
\exp (t_i\cdot\tau_j)$ go to zero.
\end{enumerate}
Assuming that Theorem 3 is proven we derive important corollary which we formulate separately for the hermitian and real symmetric cases.
\medskip

\noindent{\bf Corollary 5}
{\it
For an arbitrary Wigner hermitian matrix (1.1)--(1.3), the limit
\begin{equation}
\lim_{n\rightarrow\infty} E\ \prod^k_{i=1}\ \Trace\ A_n^{2[t_i\cdot n^{\frac{2}{3}}]+\epsilon_i},
\end{equation}
where $\epsilon_i=0,1;\ i=1,\ldots ,k$, exists for all positive $t_1,\ldots
,t_k$ and coincides with the limit in the G.U.E. case.}
\medskip

\noindent{\bf Corollary 6}
{\it
For arbitrary Wigner real symmetric matrix (1.1')--(1.3'), the limit
\begin{equation}
\lim_{n\rightarrow\infty}\ E\ \prod^k_{i=1}\ \Trace\ A_n^{2[t_i\cdot n^{\frac{2}{3}}]+\epsilon_i},
\end{equation}
where $\epsilon_i=0,1;\ i=1,\ldots ,k$, exists for all positive $t_1,\ldots
,t_k$ and coincides with the limit in the G.O.E. case.}
\medskip

\noindent{\it Remark 7}

Exact formulas for the limits in (2.19), (2.20) can be derived from
\begin{align}
\begin{split}
&\lim_{n\rightarrow\infty}\ E\ \sum_{\theta_{j_1}\neq\theta_{j_2}\neq\cdots\neq
\theta_{j_l}<n^{\frac{1}{6}}}e^{t_1\cdot\theta_{j_1}}\cdots 
e^{t_k\cdot\theta_{j_k}}\\
&=\int^\infty_{-\infty}\int^\infty_{-\infty}
e^{t_1\cdot x_1+\cdots +t_k\cdot x_k}\cdot R_{\beta ,k}(x_1,\ldots ,x_k)
dx_1\cdots dx_k.
\end{split}
\end{align}

\noindent{\it Remark 8}

A large class of linear statistics near the edge has been studied in the Gaussian case in [32].  Another work of related interest is [31], where the author studies general $\beta$ ensembles.  We also would like to draw attention to [38], [40].

The proof of Theorems 2, 3 has a strong combinatorial flavor.  Let us discuss it in more detail.  We have shown at the beginning of the section that the calculation of $E\ \Trace\ A^{p_n}_n$ can be reduced to the counting of closed paths of length $p_n$ on a nonoriented complete graph with $n$ vertices under the additional conditon that each edge will appear an even number of times.  We call the paths satisfying this condition even.  In the counting process, we assign to each path a statistical weight
$$E\ a_{i_0i_1}\cdot a_{i_1i_2}\dots a_{i_{p_n-1}i_0}.$$
We assume $p_n$ even since there are no even paths of odd length.  In the rest of the paper we shall deal with the real symmetric case; the considerations in the hermitian case are very similar.  An interesting observation can be made when $a_{ij}$ are Bernoulli random variables taking values $\pm\frac{1}{2\sqrt n}$ with probability $\frac{1}{2}$.  Then
\begin{equation}
E\ a_{i_0i_1}\cdot a_{i_1i_2}\dots a_{i_{p_n-1}i_0}=2^{-p_n}
\end{equation}
for any even path implying that $E\ \Trace\ A^{p_n}=2^{-p_n}\cdot\#\{$closed even paths of length $p_n$, with distinguished origin, on a complete nonoriented graph with $n$ vertices$\}$.

In the case of an arbitrary Wigner matrix, formula (2.22) still holds if the path has no loops (edges $\{i,i\}$) and each edge of the path appears exactly twice.  In [1], [2] Wigner proved the celebrated semicircle law by studying even paths of fixed length $p$.  He showed that the paths without loops that visit each their edge twice are typical, meaning that the ratio of the number of such paths to the number of all even paths goes to 1 as $n\rightarrow +\infty$.  To study the case when length $p_n$ is growing we need some definitions from [22], [23]:
\medskip

\noindent{\bf Definition 1}

An instant $\ell =1,2,\dots p_n-1$ is said to be marked for the closed even path $\p =\{i_0, i_1,\dots i_{p_n-1}, i_{p_n}=i_0\}$ if the nonoriented edge of $\{i_{\ell -1},i_\ell\}$ occurs an odd number of times up to the instant $\ell$ (inclusive).  The other instants are said to be unmarked.

It follows immediately from the definition that the number of marked instants for a closed even path is equal to the number of unmarked instants.
\medskip

\noindent{\bf Definition 2}

A closed even path $\p$ is called a path without self-intersections if, for any two distinct marked instants $\ell '$ and $\ell ''$, one has $i_{\ell '}\neq i_{\ell ''}$.  For purposes of Definition 2, we also assume instant 0 to be marked.

Paths without self-intersections have the following structure.  First, there is a series of marked instants when we pass through a number of distinct vertices (the number of vertices that we ``discover" during this series is equal to the length of the series).  Then there is a series of unmarked instants when we pass in the reverse order vertices visited before.  At some moment we stop the second series --- we do not necessarily ``sink" all the way down to the origin of the path --- and launch a new series of marked steps, discovering at each of the instants a new vertex.  Then we again have a series of unmarked instants when we make a few ``steps back," and so on.  One can see that for such paths, each edge appears exactly twice and there are no loops.  It is important to note that paths without self-intersections are uniquely determined by their values at marked instants.  In [1] Wigner showed that the number of closed even paths of length $p=2s$ without self-intersections is equal to $\frac{n!}{(n-s)!}\cdot\frac{(2s)!}{s!\cdot (s+1)!}$, and that such paths are typical in the limit $n\rightarrow\infty$.  It follows from this argument that the $p$th moment of the limiting spectral density of the eigenvalues is equal to 
\begin{equation}
\frac{(2s)!}{s!\cdot (s+1)!}\cdot \left (\frac{1}{4}\right )^s=\int^1_{-1} x^{2s}\cdot\frac{2}{\pi}\sqrt{1-x^2}\, dx
\end{equation}
for even $p=2s$ and 0 for odd p.  In [13] F\"uredi and Koml\'oz showed that paths without self-intersections are typical even if $p_n$ is growing not faster than $n^{\frac{1}{6}}$.  In [22] we proved that this is still true if $p_n/\sqrt{n}\underset{n\rightarrow\infty}{\longrightarrow}0$.  The values $p_n$ of order $\sqrt n$ are critical in a sense that starting with this regime, typical paths have self-intersections [23].
\medskip

\noindent{\bf Definition 3}

A marked instant $m$ is called an instant of self-intersection if there is a marked instant $m'<m$ such that $i_{m'}=i_m$.  It is important that moments $m',\ m$ in Definition 3 are required to be marked.  In Fig. 1 we give an example of a path without self-intersections.

(figure)
$$\p =\{1\rightarrow 5\rightarrow 3\rightarrow 5\rightarrow 2\rightarrow 4\rightarrow 2\rightarrow 5\rightarrow 1\}.$$
An example of a path with self-intersection is given in Fig. 2:

(figure)
$$\p =\{1\rightarrow 5\rightarrow 3\rightarrow 2\rightarrow 5\rightarrow 4\rightarrow 5\rightarrow 3\rightarrow 2\rightarrow 5\rightarrow 1\}.$$

\noindent{\bf Definition 4}

A vertex $i$ is called a vertex of simple (triple, quadruple, etc.) self-intersection if there are exactly two (three, four, etc.) marked instants m such that $i_m=i$.

It was proven in [23] that if we consider a uniform distribution on the discrete space of all closed even paths of length $p_n=2s_n$ and assume $\frac{s_n}{\sqrt n}\underset{n\rightarrow\infty}{\longrightarrow}c >0$, then the probability for nonsimple (i.e., triple, quadruple, etc.) self-intersections to occur goes to zero as $n\rightarrow\infty$, and the number of simple (or, in the light of the previous line, the number of all) self-intersections converges in distribution to Poisson law with mean $\frac{c^2}{2}$.  If $\sqrt{n}\ll p_n\ll n^{\frac{2}{3}}$ (notation ``$\ll$" means that the ratio of terms goes to $+\infty$) then the probability to have only simple self-intersections still converges to 1, and the number of simple self-intersections is
$\frac{s_n^2}{2n}(1+\underline{0}(\frac{\sqrt n}{s_n}))$ in a sense that
\begin{equation}
\frac{\#\text{ of simple self-intersections}-\frac{s_n^2}{2n}}{s_n/\sqrt{n}}
\underset{n\rightarrow\infty}{\overset{w}{\longrightarrow}}N(0,1).
\end{equation}
It is quite straightforward to derive (2.24) from results proved in [23], even though this central limit theorem was not explicitly stated there.  Among other important results, it was established that for $p_n=\bar o(n^{\frac{2}{3}})$ the probability to have some edge passed more than twice goes to zero.

To study the higher moments of Trace $A^{p_n}$, we offered a very neat approach that allows us to count $k$-tuples of closed paths of length $p_n$ satisfying an additional condition that each edge appears in the union of paths an even number of times.  In the present paper we extend these techniques to the case $p_n\sim n^{\frac{2}{3}}$.  For warm-up let us start with a more simple combinatorial problem.

\section{Toy Model}

As before, assume that we have a nonoriented complete graph $\{1,2,\dots n\}$ (i.e., every vertex is connected with any other vertex by a nonoriented edge).  In this section we shall study an ensemble of all closed paths (with distinguished origin) of length $p_n$ on the graph.  Therefore throughout this section the condition that each edge appears an even number of times no longer holds.  The number of all such paths is $n^{p_n}$ and we define a uniform distribution on the space of such paths by assigning to each path a probability $n^{-p_n}$.  First, we formulate and prove a few propositions.
\medskip

\noindent{\bf Proposition 1}
{\it
Suppose that $p_n/n\underset{n\rightarrow\infty}{\longrightarrow} 0$, then the probability of having at least one edge passed more than once goes to zero as $n\rightarrow\infty$.}
\medskip

\noindent{\bf Proof}
By a simple counting argument, the number of such paths is not greater than $\frac{p_n\cdot (p_n-1)}{2}\cdot n^2\cdot n^{p_n-4}$, and therefore the ratio of the number of such paths to the number of all paths is not greater than 
$\frac{p_n\cdot (p_n-1)}{2n^2}\underset{n\rightarrow\infty}{\longrightarrow} 0$.

Definitions 3$'$, 4$'$ are the analogues of Definitions 3, 4 in our situation.

\noindent{\bf Definition 3$'$}

An instant $m<p_n$ is called an instant of self-intersection if there exists an instant $m'<m$ such that $i_{m'}=i_m$.

\noindent{\bf Definition 4$'$}

A vertex $i$ is called a vertex of simple (triple, quadruple, etc.) self-intersection if there are exactly two (three, four, etc.) instants $m<p_n$ such that $i_m=i$.
\medskip

\noindent{\bf Proposition 2}
{\it
Let $p-N\sqrt{n}\underset{n\rightarrow\infty}{\longrightarrow} 0$.  Then the probability of having a self-intersection goes to zero.}
\medskip

\noindent{\bf Proof}
Indeed, the number of paths without self-intersections is $n\cdot (n-1)\dots
(n-p_n+1)$.  Therefore, the probability in question is $1-\prod^{p_n-1}_{k=0}(1-\frac{k}{n})$.  Since 
$\prod^{p_n-1}_{k=0}(1-\frac{k}{n})=\exp (\sum^{p_n-1}_{k=0}
\ln (1-\frac{k}{n}))=\exp (-\sum^{p_n-1}_{k=0}\frac{k}{n}-\sum^{p_n-1}_{k=0}
\frac{k^2}{2n^2}-\sum^{p_n-1}_{k=0}\frac{k^3}{3n^3}-\ldots )=\exp (-\frac{p_n(p_n-1)}{2n}+\underline{0}(\frac{p^3_n}{n^2}))$, we observe that under the condition of the proposition, $1.-\prod^{p_n-1}_{k=0}
(1-\frac{k}{n})\underset{n\rightarrow\infty}{\longrightarrow} 0.$\qed
\medskip

\noindent{\bf Proposition 3}
{\it
Let $\frac{p_n}{\sqrt n}\underset{n\rightarrow\infty}{\longrightarrow} c$.
Then the probability of having a nonsimple self-intersection goes to zero.  One can also show that the number of simple self-intersections converges in distribution to the Poisson law with mean $\frac{c^2}{2}$.}
\medskip

\noindent{\it Remark 8}

As a trivial corollary of Proposition 8, we have that the number of all self-intersections also converges to the same Poisson distribution.

\noindent{\bf Proof of Proposition 3}
We shall show that the number of paths with exactly $m$ simple self-intersections is equal to 
\begin{equation}
(c^2/2)^m\cdot \frac{1}{m!}e^{-c^2/2}\cdot n^{p_n}\cdot (1+\bar o(1)).
\end{equation}
Let us denote the instants of self-intersections by $0<j_1<j_2<\cdots <j_m<p_n$.  We are also going to use a notation $i_0=0$.  Then the number of paths that have their self-intersections (all simple) at these moments is
\begin{align}
\begin{split}
&\prod^m_{k=1}((n-j_{k-1}+k-1)\cdot (n-j_{k-1}+k-2)\dots (n-j_k+k+2)\\
&\qquad\cdot (n-j_k+k+1)\cdot (j_k-k+1))
\end{split}
\end{align}
with an agreement that if $j_k=j_{k-1}+1$, then the $k$th factor in (3.2) is $(j_k-k)$.  It is not difficult to see that the product (3.2) is equal to
\begin{align}
\begin{split}
&\prod^{p_n-1-m}_{r=0} (n-r)\cdot\prod^m_{k=1}(j_k-k+1)\cdot (1+\bar o(1))\\
&\qquad =n^{p_n}\cdot e^{-\frac{c^2}{2}}\cdot \left (\prod^m_{k=1}\frac{j_k-k+1}{n}\right )\cdot (1+\bar o(1)).
\end{split}
\end{align}
After taking the summation $0<j_1<j_2<\cdots <j_k<p_n\sim c\sqrt n$, we arrive at (3.1).  Formula (3.1) implies that the probability of having $m$ (all simple) self-intersections tends to $(c^2/2)^m\cdot\frac{1}{m}e^{-c^2/2}$.  Since limiting probabilities trivially add up to 1, the first part of Proposition 3 follows as well.\qed

One may conjecture then that triple self-intersections do not occur in typical paths which $p_n$ is of order $n^{\frac{2}{3}}$, quadruple self-intersections do not occur until $p_n$ is of order $n^{\frac{3}{4}}$, etc.  This turns out to be true.  For simplicity we consider the case $p_n=\underline{0}(n^{\frac{2}{3}})$.

\noindent{\bf Proposition 4}
Let $p_n$ go to infinity in such a way that $\frac{p_n}{\sqrt n}\underset
{n\rightarrow\infty}{\longrightarrow}+\infty$ but $\frac{p_n}{n^{\frac{2}{3}}}
\underset{n\rightarrow\infty}{\longrightarrow} 0$.  Then:
\begin{enumerate}
\item[a)] If $\eta_n$ is the number of simple self-intersections, then
$$\frac{\eta_n-\frac{p^2_n}{2n}}{p_n/\sqrt{2n}}\underset{n\rightarrow\infty}
{\overset{w}{\longrightarrow}}(0,1)$$

\item[b)] The probability of having a nonsimple self-intersection goes to zero.
\end{enumerate}

\noindent{\bf Proof}
The calculations are very similar to those in Proposition 3.  Let us denote by $Z(m)$ the number of paths with $m$ self-intersection (all simple), then the arguments above show that
\begin{equation}
Z(m)=\left (\frac{p^2_n}{2n}\right )\cdot \frac{1}{m!}e^{-p^2_n/2n}\cdot
n^{p_n}\cdot (1+\bar o(1))
\end{equation}
uniformly in $0\leq m\leq 10\frac{p^2_n}{n}$ (we can replace 10 here by any other constant).  Then b) can be proved as before and a) follows from the Central Limit Theorem for a sum of independent identically distributed Poisson random variables.

\noindent{\bf Proposition 5}
Let $p_n/n^{\frac{2}{3}}\underset{n\rightarrow\infty}{\longrightarrow} c$.  Then
\begin{enumerate}
\item[a)] The probability of having a self-intersection of any kind other than simple or triple goes to zero as $n\rightarrow\infty$.

\item[b)] The number of triple self-intersections converges in distribution to Poisson law with mean $\frac{c^3}{6}$.
\end{enumerate}

\noindent{\bf Proof}
This is the first time in this section when some technicalities may appear.  Trying to imitate the proofs of Propostions 3, 4 we denote by
\begin{equation}
0<j^{(1)}_1<j^{(1)}_2<\cdots <j^{(1)}_m<p_n
\end{equation}
the instants of $m$ simple self-intersections, and by
\begin{align}
\begin{split}
&(j^{(2)}_{1,1},\ j^{(2)}_{1,2}),\ (j^{(2)}_{2,1}, j^{(2)}_{2,2}),\ldots
,(j^{(2)}_{\ell ,1},\ j^{(2)}_{\ell ,2}),\\
&\qquad
0< j^{(2)}_{1,1}<j^{(2)}_{2,1}<j^{(2)}_{3,1}<\cdots <j^{(2)}_{\ell ,1}<p_n,\\
&\qquad 0<j^{(2)}_{k,1}<j^{(2)}_{k,2}<p_n\ ,\ k=1,2,\ldots ,\ell
\end{split}
\end{align}
the pairs of instants corresponding to $\ell$ triple self-intersections.  The notations mean that we revisit a site of $r$th simple self-intersection, $r=1,
\ldots ,m$ at the instant $j^{(1)}_r$, and a site of $k$th triple self-intersections, $k=1,\ldots ,\ell$, at the instants $j^{(2)}_{k,1}<j^{(2)}_{k,2}$.  Let us denote by $T(\ell )$ the number of paths that have exactly $\ell$ triple self-intersections and no self-intersections of higher order.  Then employing a counting argument, one may hope to end up with something like this:
\begin{align}
\begin{split}
&T(\ell )\sim\sum^{(p_n-3\ell )/2}_{m=0}\sum_j\left (\prod^{p_n-1-m-2\ell}_{t=0}(n-t)\right )\\
&\qquad
\cdot\prod^m_{r=1}(j^{(1)}_r-r+1)\cdot\prod^\ell_{k=1}(j^{(2)}_{k,1}-k+1)
\cdot (1+\bar o(1)),
\end{split}
\end{align}
where the sum $\sum_j$ is taken over indices (3.5), (3.6). 
However, one promptly realizes that arguments similar to those from Proposition 3 would just prove that the r.h.s.~of (3.7) is an upper bound of $T(\ell )$.
Therefore slightly different arguments are needed to finish the proof.
Actually the proof we offer below is easier than the outlined approach.  Yet
we spend some time discussing it on purpose since some reincarnation of these arguments will be used in \S 4 to derive estimates from above for $E\ \Trace\ 
A^{2\cdot[t\cdot n^{\frac{2}{3}}]}$ and higher moments.  To proceed with the proof of the proposition, we observe that the number of closed paths with $m$ simple self-intersections, $\ell$ triple self-intersections and zero 
higher-order self-intersections is equal to
\begin{align}
\begin{split}
&\frac{p_n!}{(p_n-2m-3\ell )!(2m)!(3\ell )!}\cdot\frac{(2m)!}{(m!)\cdot (2!)^m}\cdot
\frac{(3\ell )!}{\ell !\cdot (3!)^\ell}\\
&\qquad\cdot\frac{n!}{(n-p_n+m+2\ell )!}
\end{split}
\end{align}
Let us assume for a minute that
\begin{equation}
\left\vert m-\frac{p^2_n}{2n}\right\vert <n^{\frac{1}{4}},\ \ell <\log n.
\end{equation}
By Stirling's formula, (3.8) is equal to
\begin{align}
\begin{split}
&\frac{p_n^{p_n}}{e^{p_n}}\cdot\sqrt{2\pi\cdot p_n}\cdot
\frac{e^{p_n-2m-3\ell}}{(p_n-2m-3\ell )^{p_n-2m-3\ell}}\cdot\frac{1}
{\sqrt{2\pi \cdot (p_n-2m-3\ell )}}\\
&\qquad\cdot\frac{1}{2^m\cdot m!}\cdot\frac{1}{6^\ell\cdot\ell !}\cdot
\prod^{p_n-2m-3\ell}_{t=0} (n-t).
\end{split}
\end{align}
Writing 
\begin{align*}
\begin{split}
\prod^{p_n-1-m-2\ell}_{t=0}(n-t)&=n^{p_n-m-2\ell}
\cdot\exp (-\frac{(p_n-1-m-2\ell )(p_n-m-2\ell )}{2n}\\
&-\frac{1}{6n^2}\cdot (p_n-1-m-2\ell )^3 )\cdot (1+\bar o(1))\\
&=n^{p_n-m-2\ell}\cdot\exp \left (-\frac{p_n^2}{2n}+\frac{p_n^3}{2n^2}-\frac{p_n^3}{6n^2}\right )\cdot (1+\bar o(1))
\end{split}
\end{align*}
and 
\begin{align*}
\begin{split}
\left (\frac{p_n}{(p_n-2m-3\ell )}\right )^{p_n-2m-3\ell}&=
\left (1+\frac{2m+3\ell}{p_n-2m-3\ell}\right )^{p_n-2m-3\ell}\\
&=\exp (2m+3\ell -((2m+3\ell )^2/(2p_n-4m-6\ell )))\\
&\qquad\cdot (1+\bar o(1))=\exp (2m+3\ell -\frac{p_n^3}{2n^2})
\cdot (1+\bar o(1)),
\end{split}
\end{align*}
 we conclude that the 
probability of having exactly $m+\ell$ self-intersections, $m$ of which are simple and $\ell$ are triple, is
\begin{equation}
\left (\frac{p_n^2}{2n}\right )\cdot\frac{1}{m!}e^{-\frac{p_n^2}{2n}}
\cdot\left (\frac{p_n^3}{6n^2}\right )^\ell\cdot\frac{1}{\ell !}\cdot
\ell^{\frac{p_n^3}{6n^2}}\cdot (1+\bar o(1))
\end{equation}
uniformly in $\vert m-\frac{p_n^2}{2n}\vert <n^{\frac{1}{4}},\ \ell <\log n$.
One readily recognizes (2.31) as a two-dimensional Poisson distribution.  Then
$$\sum_{m:\vert m-\frac{p_n^2}{2n}\vert <n^{\frac{1}{4}}}\left (\frac{p_n^2}{2n}\right )^m\cdot\frac{1}{m!}e^{-\frac{p_n^2}{2n}}=1-\bar o(1),$$
which implies that the probability of having $\ell$ triple self-intersection
(together with some number of simple self-intersections) is equal to
$(\frac{c^3}{6})^\ell\cdot\frac{1}{\ell !}e^{-\frac{c^3}{6}}\cdot
(1+\bar o(1))$.\qed

A generalization of Propositions 4, 5 is quite straightforward.

\noindent{\bf Proposition 6}
{\it
\begin{enumerate}
\item[a)] Let $p_n=\bar o(n^{\frac{k-1}{k}}),\ k=2,3,\dots$.  Then the probability of having a self-intersection of order $k$ or higher is going to zero as $n\rightarrow\infty$.

\item[b)] Let the limit of the ratio $p_n/n^{\frac{k-1}{k}}$ exist and equal to $c$.  Then the number of self-intersection of order $k$ is distributed in the limit according to Poisson law with the mean $\frac{c^k}{k!}$.
\end{enumerate}
}

In [22], [23] we showed that the analogues of Propositions 1--4 hold when we impose an additional condition for closed paths to be even.  One can view Theorem 2 in this paper as a partial result toward establishing Proposition 5 for even closed paths.  These results suggest that the following conjecture may be true.  Consider an ensemble of closed paths of length $p_n=m\cdot s_n$
on the complete nonoriented graph with $n$ vertices, with an additional condition that each edge appears in the path a number of times divided by $m$ (if $m=2$ such paths are exactly even paths ).  Definitions 2--4 then need trivial modifications.  We shall formulate our conjecture as an open problem.
\medskip

\noindent{\bf Open Problem}:  To prove the analogues of Propositions 1--6.

\section{Proof of Theorems 2 and 3}

The considerations in this section are very close to those in \S 4,5 from [23].  We shall start with Theorem 2.  Since for odd $p_n\ E\ \Trace\ 
A^{p_n}_n=0$, we assume $p_n$ even, $p_n=2s_n$.  According to Definition 4 from \S 2 all the vertices split into $s_n+1$ disjoint subsets:
$\{1,\dots n\}=\bigsqcup^{s_n}_{k=0}\n_k$ with respect to the path $\p$, where $\n_k$ is the subset of vertices of $k$-fold self-intersections.  All vertices of $\n_0$ with possible exception of the initial point of the path $i_0$ do not belong to $\p$.  Denoting $n_k=\#(\n_k)$ we see that $\sum^{s_n}_{k=0}n_k=n$,
\begin{equation}
\sum^{s_n}_{k=0}k\cdot n_k=s_n.
\end{equation}
We say the $\p$ is a path of type $(n_0, n_1, \ldots ,n_{s_n})$.  It is easy to see that every path without self-intersections is a path of type $(n-s_n,
s_n, 0,0,\ldots ,0)$.  In the condition of Theorem 2, we assume that $s_n=[t\cdot n^{\frac{2}{3}}],\ t>0$.  It will then follow from our proof that the probability of having $n_j=0$ for all $j>3$ goes to 1.  (If Proposition 6 from the previous section also holds for even closed paths, it will imply that under the condition $s_n=\bar o(n^{\frac{k-1}{k}})$ for typical paths, $n_j=0$ for
$j\geq k$, and if $s_n\sim n^{\frac{k-1}{k}}$, then for typical paths $n_k$ is of order of constant.)

In [23] we introduced the notions of closed and nonclosed vertices of simple self-intersections and proved that for $s_n=\bar o
(n^{\frac{2}{3}})$, all vertices of simple self-intersections for typical paths are closed.  Below we construct examples of closed and nonclosed vertices of simple self-intersections.

(figure)
$$\p =\{i\rightarrow j\rightarrow k\rightarrow\ell\rightarrow k\rightarrow j\rightarrow m\rightarrow k\rightarrow n\rightarrow k\rightarrow m\rightarrow j\rightarrow i\}.$$
For such path $i\in\n_0;\ j,\ell ,m, n\in\n_1;\ k\in\n_2$, and $k$ is a closed vertex of simple self-intersection.  Note that $j$ belongs to $\n_1$, not to $\n_2$, since we arrive at $j$ at the marked instant only once (from $i$), while during three other arrivals at $j$ we pass through corresponding edges for a second time.  The vertex $k$ in this example is closed because if, after the moment of self-intersection, we wanted to leave $k$ along the already appeared edge, we had only one such possibility (along the edge $\{k\ m\}$).

(figure)
$$\p =\{i\rightarrow j\rightarrow k\rightarrow m\rightarrow j\rightarrow q\rightarrow j\rightarrow k\rightarrow m\rightarrow j\rightarrow i\}.$$
For this path $i_0\in\n_0;\ k,m,q\in\n_1;\ j\in\n_2$, and $j$ is a nonclosed vertex of simple self-intersection, because if we wanted to leave $j$ after the moment of self-intersection along the already appeared edge, we had more than one such opportunity $(\{j,\ k\},\ \{j,\ m\},\} \{j,\ i\}$; in the case of $\p$ we choose $\{j,\ k\})$.  As we already explained before, even closed paths without self-intersections possess a remarkable property; the trajectory of the pass is determined uniquely by the initial point and the restriction of the trajectory to the marked instants.  This property simplifies a great deal the problem of counting such paths.  For a path with self-intersections, the choice of continuations of trajectory during unmarked instants (the choice of the ``backward trajectory") may not be unique (see Example 2).  For example, for the ``first return" from a vertex of simple self-intersection, one of the following three edges can be chosen:
\begin{enumerate}
\item[(a)] the edge we used to arrive at the vertex for the first time,

\item[(b)] the edge we used to leave the vertex right after the first arrival,

\item[(c)] the edge we used to arrive at the vertex for the second time.
\end{enumerate}
One can see that in Example 2 we have chosen possibility (b).
Such considerations prompted us to call a vertex of simple self-intersection
closed if there is a unique way of continuing the trajectory at an unmarked step when we ``return" from the vertex.  Otherwise, we call a vertex nonclosed.  A crucial observation made in [23] is that probability for a vertex of simple self-intersection to be nonclosed is of order of $\frac{1}{\sqrt {s_n}}$.  Since the number of all self-intersections is $\underline 0(\frac{s_n^2}{n})$ we see then that for typical paths, there are no nonclosed vertices provided $\frac{s_n^{\frac{3}{2}}}{n}\underset{n\rightarrow\infty}{\longrightarrow} 0.$
We will show below that if $s_n\sim n^{\frac{2}{3}}$, then for typical paths the number of nonclosed vertices of simple self-intersections is of order of constant.

Let us start with the formula (2.1):
$$E\ \Trace\ A_n^{p_n}=\sum^n_{i_0,i_1,\cdots i_{p_n-1}=1}E\ a_{i_0,i_1}
a_{i_1,i_2}\dots a_{i_{p_n-1},i_0}.$$
We have shown in [23] that a subsum of (2.1) over the paths of type $(n_0,n_1,
\ldots ,n_{s_n})$ is bounded from above by
\begin{align}
\begin{split}
&n^{-s_n}\cdot\frac{n!}{n_0!n_1!\dots n_{s_n}!}n\cdot\frac{(2s_n)!}{s_n!\cdot
(s_n+1)!}\cdot\frac{s_n!}{\prod^{s_n}_{k=1}(k!)^{n_k}}\cdot\\
&\qquad\cdot 4^{-s_n}\cdot\prod^{s_n}_{k=2}(\const_1\cdot k)^{2k\cdot n_k}
\end{split}
\end{align}
The last inequality followed from
\begin{align}
\begin{split}
&\underset{(n_0,n_1,\dots n_{s_n})}{\underset{\p\text{of type}}{\text{max}}}
E \prod^{2s_n-1}_{\ell =0}\xi_{i_\ell i_{\ell +1}}\cdot W_n(\p\mid
\text{marked instants})\\
&\qquad\leq (4n)^{-s_n}\cdot 3^r\cdot\prod^{s_n}_{k=3}(\const\ k)^{kn_k},\ 
\const>0,
\end{split}
\end{align}
where $W_n$ is the number of ways the trajectory can be chosen at unmarked instants
provided that the vertices at the marked instants have already been chosen, and $r$ is the number of nonclosed vertices.  One can show then the existence of another positive constant const$_2>0$ such that the subsum of (2.1)
over the paths for which $\sum^{s_n}_{k=2}k\cdot n_k\geq
\const_2\cdot\frac{s_n^2}{n}$ tends to zero as $n\rightarrow\infty$.
The actual value of const$_2$ is not important. One can show for example, that const$_2=10$ is enough.  Therefore we restrict our attention to the paths for which
\begin{equation}
\sum^{s_n}_{k=2}k\cdot n_k<10\frac{s^2_n}{n}.
\end{equation}
Our counting strategy will be the following.  First we associate to every path $\p$ a trajectory $X=\{x(t),\ 0\leq t\leq 2s_n\}$ of a simple walk on the nonnegative half-lattice.  The trajectory starts and ends at zero, $x(0)=x(2s_n)=0$, and if $0<t_1<t_2<\cdots <t_{s_n}<2s_n$ are the marked instants, then $x(t)-x(t-1)=1$ if $t$ is marked, and $x(t)-x(t-1)=-1$ if $t$ is unmarked.  Let $\p$ have $M$ instants of self-intersection, that is, $ M=\sum^{s_n}_{k=2} (k-1)\cdot n_k$.  We denote by
\begin{equation}
t_{j^{(1)}_1}<t_{j^{(1)}_2}<\cdots <t_{j^{(1)}_{n_2}},\ 1\leq j^{(1)}_1<
j^{(1)}_2<\cdots < j^{(1)}_{n_2}\leq s_n
\end{equation}
the instants of simple self-intersections.  Some of these instants may correspond to nonclosed vertices.  Assume that the number of nonclosed vertices is $r$, and the instants of simple self-intersections corresponding to the nonclosed vertices are
\begin{equation}
t_{u_1}<t_{u_2}<\cdots <t_{u_r},\ 1\leq u_1 < u_2 <\cdots < u_r\leq s_n.
\end{equation}
We denote the pairs of indices of marked instants corresponding to triple self-intersections by $(j^{(2)}_{1,1},\ j^{(2)}_{1,2}),\ (j^{(2)}_{2,1},\ 
j^{(2)}_{2,2}),\ldots , (j^{(2)}_{n_3,1},\ j^{(2)}_{n_3,2})$ and order them:
\begin{equation}
1\leq j^{(2)}_{1,1}<j^{(2)}_{2,1}<\cdots < j^{(2)}_{n_3,1}\leq s_n,\ \ 
j^{(2)}_{\ell ,1} < j^{(2)}_{\ell, 1},\ \ell =1,\dots , n_3.
\end{equation}
Notations in (4.7) mean that the $\ell$th vertex of triple self-intersection is visited for a second time at a marked instant $t_{j^{(2)}_{\ell , 1}}$, and for a third time at a marked instant $t_{j^{(2)}_{\ell , 2}}$.  Similarly, we denote by $(j^{(2)}_{1,1} ,\ldots , j^{(k)}_{1,k}),\dots , (j^{(k)}_{n_{k+1} ,1},\dots , j^{(k)}_{n_{k+1} ,1})$ the $k$-tuples of indices of $(k+1)$-fold self-intersections.  We order them in such a way that
\begin{align}
\begin{split}
1\leq j^{(k)}_{1 ,1}<j^{(k)}_{2 ,1}&<\cdots < j^{(k)}_{n_{k+1} ,1}\leq s_n,\\ 
j^{(k)}_{\ell ,1}< j^{(k)}_{\ell ,2}&<\cdots <j^{(k)}_{\ell, k}, \ \ \ell =1,\ldots , n_{k+1}.
\end{split}
\end{align}
The notations imply that the $\ell$th vertix of $(k+1)$-fold self-intersection is visited at marked instants $t_{j^{(k)}_{\ell , 1}},\ldots , t_{j^{(k)}_{\ell , k}}$ after the first visit has occurred.

To refine our classification of simple self-intersections, we shall do the following.  If we look at two edges along which we arrived at marked instants at a vertex of simple self-intersection, then there are two possibilities depending on whether such two edges coincide or not.  If they coincide, then the edge appears in the path four times (twice at marked instants and twice at unmarked instants).  Let us denote the number of vertices from class $\n_2$
which we visit both times at marked instants along the same edge by $q$, and the corresponding instants by
\begin{equation}
t_{v_1}<t_{v_2}<\cdots <t_{v_q},\ \ 1\leq v_1<v_2<\cdots <v_q\leq s_n.
\end{equation}
Finally we introduce the following characteristic of $\p$: the maximum of all numbers of vertices that can be visited at marked instants from a vertex of the path.  We denote this maximum by $\nu_n(\p )$.  By definition, each vertex of the path can be the left end of at most $\nu_n(\p )$ marked edges.  The actual order of $\nu_n (\p )$ for typical even paths is not known (it is probably $\log s_n$).  We shall show below that the subsum of (2.1) over paths with $\nu_n(\p )>s_n^{\frac{1}{2}-\epsilon}$ is $\bar o(1)$ (actually we can replace $\frac{1}{2}-\epsilon$ in the exponent by any $\gamma >0$).
Let us denote by $Z(n_1,n_2,\ldots ,n_{s_n};\ r,q)$ a subsum of (2.1) over the paths with fixed $n_k,\ k=1,\ldots ,s_n$, and $r,q$, that also satisfy the condition
$\sum^{s_n}_{k=2}k\cdot n_k<10\frac{s^2_n}{n}.$
We shall split it into two sums:
$$Z(n_1,n_2,\ldots ,n_{s_n};\ r,q)=Z'(n_1,n_2,\ldots ,n_{s_n};\ r,q)+
Z''(n_1,n_2,\ldots ,n_{s_n};\ r,q),$$
where the first subsum is over the paths for which $\nu_n(\p )\leq
s_n^{\frac{1}{2}-\epsilon},\ \epsilon$ fixed, and the second is over the rest.
We define $Z'$ as a sum of all $Z'(n_1,\ldots ,n_{s_n};\ r, q)$, and similarly for $Z''$.  Our next goal is to obtain a nice (Poisson) upper bound for $Z'(n_1,n_2,\cdots ,n_{s_n};\ r,q)$.  Let $\Omega$ be a collection of all
$X=\{x(t)$, $t=0,1,\dots ,2s_n\}$ for which $x(0)=x(2s_n);\ x(t)\geq 0;\ x(t+1)-x(t)=\pm 1\ \ \forall t$.  The number of elements in $\Omega$ is equal to $\frac{(2s_n)!}{s_n!\cdot (s_n+1)!}^{-1}$.  The mathematical expectation with respect to such probability distribution on $\Omega$ will be denoted by $E_{n,X}$.
\medskip

\noindent{\bf Lemma 5}
{\it
There are positive constants $A, B, C, D$ such that
\begin{align}
\begin{split}
&Z'(n_1,\ldots ,n_{s_n},r,q)\leq n\cdot\frac{(2s_n)!}{s_n!\cdot (s_n+1)!}\cdot 4^{-s_n}\\
&\qquad\cdot\frac{1}{(n_x-r-q)!}\cdot\left (\frac{s^2_n}{2n}\right )^{n_2-r-q}
\cdot e^{-\frac{s^2_n}{2n}}\cdot e^{A\cdot\frac{s^3_n}{n^2}}\\
&\qquad
\cdot\frac{1}{r!}\cdot\left (\frac{B\cdot s^{\frac{3}{2}}_n}{n}\right )^r
\cdot \left (E_{n,x}(\underset{1\leq t\leq 2s_n}{\max}\frac{x(t)}{\sqrt {s_n}})^r\right )\\
&\qquad\cdot\frac{1}{q!}\cdot\left (\frac{c\cdot s^{\frac{3}{2}-\epsilon}_n}
{n}\right )^q\cdot\frac{1}{n_3!}\cdot\left (\frac{(D\cdot s_n)^3}{n^2}
\right )^{n_3}\\
&\qquad\cdot\prod^{s_n}_{k=4}\frac{1}{n_4!}\left (\frac{(D\cdot s_n)^k}
{n^{k-1}}\right )^{n_k}.
\end{split}
\end{align}
}

\noindent{\bf Proof}
Define
\begin{align}
\begin{split}
&\{j_1,j_2,\cdots ,j_{n_2-r-q}\}=\{j^{(1)}_1,\ldots ,j^{(1)}_{n_2}\}\setminus (\{u_1,\ldots
,u_r\}\cup\{v_1,\cdots ,v_q\}),\\
&\qquad 1\leq j_1<j_2<\cdots <j_{n_2-r-q}\leq s_n.
\end{split}
\end{align}
By a simple counting argument we have 
\begin{align}
\begin{split}
&Z'(n_1,n_2,\ldots ,n_{s_n},r,q)\leq\sum_{X\in\Omega}\ \ 
\sum_{\stackrel{\text{over indices}}{\text{in (4.11)}}}\ \ 
\sum_{\stackrel{\text{over indices}}{\text{in (4.6)}}}\ \  
\sum_{\stackrel{\text{over indices}}{\text{in (4.9)}}}\ \ 
\sum_{\stackrel{\text{over indices}}{\text{in (4.7)}}}\\
&\qquad\sum^{s_n}_{k=3} \ \
\sum_{\stackrel{\text{over indices}}{\text{in (4.8k)}}}
\prod^{n_1+\cdots +n_{s_n}}_{y_1=0}(n-y_1)\cdot \prod^{n_2-r-q}_{y_2=1} j_{y_2}\cdot \prod^r_{d=1} x(t_{u_d})\cdot\prod^q_{\ell =1}
s_n^{\frac{1}{2}-\epsilon}\cdot\\
&\qquad\prod^{n_3}_{y_3=1}j^{(3)}_{y_3,1}\cdot
\prod^{s_n}_{k=4}\prod^{n_k}_{y_k=1}
j^{(k)}_{y_k,1}\cdot
\frac{1}{n^{s_n}}\cdot\frac{1}{4^{s_n}}\cdot 3^r \cdot (\const\cdot 2)^{2q}\cdot\prod^{s_n}_{k=3}(\const\cdot k)^{k\cdot n_k}
\end{split}
\end{align}
(if $n_k=0$ for some $k$, we assume the corresponding factor is one).

We hope that the reader is not scared by the array of sums and products in the formula.  Actually it is quite self-explanatory:
\begin{enumerate}
\item[a)] Each trajectory of a simple walk $X$ leaves us with a choice of marked instants.

\item[b)] The product $\prod^{n_1+\cdots n_{s_n}}_{y_1=0} (n-y_1)$ gives us the number of possibilities for choosing all the vertices that will appear in the path in the order of their appearance.

\item[c)] Since some vertices may appear more than once, the choice of indices in (4.11), (4.6), (4.9), (4.7), (4.8) lets us set up the moments of self-intersections.

\item[d)] The product
$$\prod^{n_2-r-q}_{y_2=1}j_{y_2}\cdot\prod^r_{d=1} x(t_{u_d})\cdot\prod^q_{\ell =1}(s_n^{\frac{1}{2}-\epsilon})$$
gives us an estimate from above for choosing the vertices of simple self-intersections.  Indeed, at any moment $t_{j^{(1)}_\alpha}$ there are no more than $j^{(1)}_\alpha$ possibilities for choosing a vertex (that will be a vertex of simple self-intersection) among previously appeared by this moment vertices.  If $j^{(1)}_\alpha$ is from (4.6) then we have to pick a nonclosed vertex, therefore the number of possibilities is even smaller.  One can see that this may be done in no more than $x(t_{j^{(1)}_\alpha})$ ways.  Finally if $j^{(1)}_\alpha$ is from (4.9), then we have to take the preceding vertex in the path and choose from among all vertices connected to that one by an edge from the path.  Then we have no more than $\nu_n(\p )\leq s^{\frac{1}{2}-\epsilon}_n$ possibilities.

\item[e)]  Similar arguments apply to the products over $y_k,\ k=3,\dots
s_n$, when we are choosing the vertices of self-intersections belonging to 
$\n_k$.  The choices made in a)--e) let us uniquely determine the restriction of $\p$ to the set of marked instants and the initial point.  Therefore we are left with the problem of estimating $ E\ \prod^{2s_n-1}_{d=0}
\xi_{i_di_{d+1}}$ and $W_n(\p\mid$ marked instants) --- the number of possible choices for continuing the path at unmarked instants provided the vertices at marked instants are known.  It immediately follows from $(1.3),\ 
(1.3')$ that
\begin{equation}
E\ \prod^{2s_n-1}_{d=0}\xi_{i_di_{d+1}}\leq n^{-s_n}\cdot\prod^{s_n}_{k=1}
(\const\cdot k)^{k\cdot n_k}.
\end{equation}
As for $W_n$ one can write 
\begin{equation}
W_n\leq\prod^{s_n}_{k=1} (2k)^{k\cdot n_k},
\end{equation}
arguing that at each unmarked instant of ``return" from a vertex of $k$-fold self-intersection, we have at most $2k$ possibilities to choose the next vertex.
The last two inequalities give
$$\left (E\ \prod^{2s_n-1}_{d=0}\xi_{i_di_{d+1}}\right )
\cdot W_n\leq n^{-s_n}\cdot\prod^{s_n}_{k=1}
(2\ \const\cdot k^2)^{k\cdot n_k}.$$
\end{enumerate}
It appears that one can prove a better estimate:
\begin{align}
\begin{split}
&\left (E\ \prod^{2s_n-1}_{d=0}\xi_{i_di_{d+1}}\right )
\cdot W_n\leq\frac{1}{n^{s_n}}
\cdot \frac{1}{4^{s_n}}\cdot 3^r\\
&\qquad\cdot (\const\cdot 2)^{2\cdot q}\cdot
\prod^{s_n}_{k=3}(\const_1\cdot k)^{k\cdot n_k}.
\end{split}
\end{align}
The idea behind (4.15) is that the two factors in the l.h.s. of (4.15) cannot be simultaneously too big --- if some edge appears in $\p$ a large number of times increasing the first factor, then we will use this edge for ``return" many times, thus decreasing the number of possible continuations of the trajectory at the unmarked instants.

Formula (4.15) was proven in Lemma 1 [23].  (We proved it there for $q=0$ and the argument can be trivially generalized for any $q$.  Looking at the original proof in [23], one can also notice a typographical error --- the factor $\frac{1}{n^s}$ is missing there.)

Once (4.12) is proven, the result of Lemma 5 can be proven by taking a summation there.  The considerations follow closely those from [23], \S 4, and actually are not very difficult.  As a corollary of (4.10), we have
\medskip

\noindent{\bf Lemma 6}
{\it
Let $s_n$ grow to infinity such that $s_n=\underline{0}(n^{\frac{2}{3}})$.  Then
\begin{align}
\begin{split}
Z'&\leq\frac{1}{\pi}\cdot\frac{n}{s^{\frac{3}{2}}_n}\cdot\exp (A\cdot\frac{s^3_n}{n^2})\cdot E_{n,X}\exp \left (B\cdot\frac{s_n^{\frac{3}{2}}}{n}\cdot\underset{[0,2s_n]}{\max}
\frac{x(t)}{\sqrt {s_n}}\right )\\
&\leq \frac{1}{\pi}\cdot\frac{n}{s_n^{\frac{3}{2}}}\cdot\exp\left (\gamma\cdot
\frac{s_n^3}{n^2}\right )
\end{split}
\end{align}
with some positive constant $\gamma$.}
\medskip

\noindent{\bf Proof}
The first inequality follows from (4.10) by summation.  The second one follows
from the fact that the tail of the distribution of the normalized maximum
decays as fast as Gaussian uniformly in $n$ (which is a nice exercise).

Our next step is to show that $s_n$ proportional to $n^{\frac{2}{3}}$, the second subsum of (2.1), $Z''$, vanishes in the limit $n\rightarrow\infty$.
Let us denote by $Z'' (n_1,n_2,\ldots ,n_{s_n},r,\break\nu_n)$ a subsum of (2.1) with fixed 
$n_1,\cdots ,n_{s_n},r,\nu_n$ and assume $\nu_n>s_n^{\frac{1}{2}-\epsilon}$.  By definition, the sum over all such subsums is $Z''$.  To formulate the analogue of Lemma 5, we need some more notations.  Let
$ N=r +\sum^{s_n}_{k=3} k\cdot n_k$, and 
\begin{equation}
0\leq t_1<t_2<\cdots t_n\leq 2s_n
\end{equation}
be some integers.  We denote by $\Gamma_{t_1,\dots t_N}$ an event from
$\Omega$ such that $\Gamma_{t_1,\dots t_N}$ consists of trajectories of simple walks $X$ for which the following holds:  There is an interval among
$[t_i,t_{i+1}],\ i=1,\dots N$, such that the trajectory of $X$ restricted to a subinterval of the interval descends to some level at least $[\frac{\nu_n}{N}]$ times but never crosses it (see fig. 5):

(figure)
\medskip

\noindent{\bf Lemma 7}
{\it
There are positive constants $A, B,C,D$ such that
\begin{align}
\begin{split}
&Z'' (n_1,\ldots n_{s_n},r,q,\nu_n)\leq n\cdot\frac{(2s_n)!}{s_n!\cdot
(s_n+1)!}\cdot 4^{-s_n}\\
&\qquad
\cdot\frac{1}{(n_2-r-q)!}\cdot\left (\frac{s^2_n}{2n}\right )^{n_2-r-q}\cdot e^{-\frac{s^2_n}{2n}}\cdot e^{A\cdot\frac{s^3_n}{n^2}}\cdot\\
&\qquad
\cdot\frac{1}{r!}\cdot\left (\frac{Bs_n^{\frac{3}{2}}}{n}\right )^r\cdot\frac
{1}{q!}\cdot\left (\frac{C\cdot s_n\cdot\nu_n}{n}\right )^q\cdot\\
&\qquad
\cdot\frac{1}{n_3!}\cdot\left (\frac{(D\cdot s_n)^3}{n^2}\right )^{n_3}
\cdot\prod^{s_n}_{k=4}\frac{1}{n_k!}\left (\frac{(D\cdot s_n)^k}
{n^{k-1}}\right )^{n_k}\cdot\\
&\qquad
\cdot\underset{t_1<\cdots <t_N}{\max}\left (E_{n,X}\ \left (\max\frac{x(t)}
{\sqrt {s_n}}\right )^\Gamma\cdot \chi_{\Gamma_{t_1,\dots t_N}}\right ),
\end{split}
\end{align}
where $\chi_{\Gamma_{t_1\ldots t_N}}$ is an indicator of the set $\Gamma_{t_1
\ldots t_N}$.}
\medskip

\noindent{\bf Proof}
The proof becomes very much similar to that of Lemma 5 once we realize that fixing the value of $\nu_n$ translates into $X\in\Gamma_{t_1,\dots t_N}$ with $t$'s being the instants of $k$-fold self-intersections, $k\geq 3$,
and nonclosed simple self-intersections.  Indeed, if a vertex is the left end of $\nu_n$ marked edges of a paths with simple self-intersections all of which correspond to closed vertices, then for the corresponding random walk $X=\{x(t), 0\leq t\leq 2s_n\}$ there exists a time interval $[T_1,T_2]$
on which trajectory descends $\nu_n$ time to the level $x(T_1)$ but never crosses it (i.e., never descends to the level $x(T_1)-1)$.  This is exactly because in order to make each new step from the vertex $i_{T_1}$ we must first return to it along the path.  Generally at least one of the intervals 
$[t_1,t_2],\dots ,[t_{N-1},t_N]$ will have $[\frac{\nu_n}{N}]$ of such returns, which completes the argument.\qed

Let us denote now by $Z''(N,\nu_n)$ the sum of $Z''(n_1,\ldots ,n_{s_n},
r,q,\nu_n)$ over $r, q$ and $n_1,\ldots n_{s_n}$ such that 
$ r+\sum^{s_n}_{k=3}n_k=N$ is fixed.  As a corollary of (4.17), we have 
\begin{align}
\begin{split}
&Z''(N,\nu_n)\leq n\cdot\frac{(2s_n)!}{s_n!\cdot (s_n+1)!}\cdot 4^{-s_n}\cdot
e^{A\cdot\frac{s^3_n}{n^2}}\cdot e^{C\cdot\frac{s_n\cdot\nu_n}{n}}\cdot \frac{1}{N!}\\
&\qquad\left (\frac{\const\cdot s^3_n}{n^2}\right )^N\cdot
\underset{t_1<\cdots <t_N}{\max}\ E_{n,x} \biggl (\exp \left (B
\frac{s_n^{\frac{3}{2}}}{n}\cdot \max\frac{x(t)}{\sqrt{s_n}}\right )
\cdot \chi_{\Gamma_{t_1,\cdots t_N}}\biggr ).
\end{split}
\end{align}
It is an exercise to show that the probability of $\Gamma_{t_1,\cdots ,t_N}$
is exponentially small in 
\begin{equation}
\nu_n/N:\ P(\Gamma_{t_1,\cdots ,t_N})\leq (2s_n)^2
e^{-\const\frac{\nu_n}{n}},
\end{equation}
which is intuitively clear since at each of $[\frac{\nu_n}{N}]$ times, the next step is predetermined (we have to go north).  Equations (4.19) and (4.20) imply
\begin{align}
\begin{split}
&Z''=\sum^{s_n}_{N=0}\ \ \sum^{s_n}_{\nu_n=s_n^{\frac{1}{2}-\epsilon}+1}\ Z''
(N,\nu_n)\leq\sum^{s_n}_{N=0}\ \ \sum^{s_n}_{\nu_n=s_n^{\frac{1}{2}-\epsilon}+1}\frac{1}{\pi}\frac{n}{s_n^{\frac{3}2}}\cdot\\
&\qquad e^{\const\ s_n^3/n^2}\cdot
\frac{1}{N!}\left (\frac{\const\cdot s_n^3}{n^2}\right )^N\cdot 
e^{(C\cdot s_n\cdot\nu_n)n}\cdot (2s_n)^2\cdot e^{-\const\cdot\nu_n/N}
\end{split}
\end{align}
(we remind that the notation const is used for different constants throughout the paper).  Finally one can show that $s_n\sim n^{\frac{2}{3}}$ and (4.20) imply
\begin{equation}
Z''=\bar o(1).
\end{equation}
This finishes the proof of part a) of Theorem 2.  Formulas (4.10), (4.21) also imply part b) since the sum over $q+\sum^{s_n}_{k=4}n_k>0$ is clearly $\bar o(1)$, and one can proceed in a similar way to show that for typical paths, there are no loops and among a finite number of vertices from $\n_3$ class, none has a corresponding edge to appear more than twice.
\qed

In the previous two papers we explained an idea that allows us to deal with
higher moments in a similar manner.  We start with an obvious formula
\begin{align}
\begin{split}
&E\ \prod^k_{m=1} (\Trace\ A^{p^{(m)}_n}_n-E\ \Trace\ A^{p^{(m)}_n}_n)=
n^{-(p^{(1)}_n+\ldots +p^{(k)}_n)/2}\cdot \\ 
&E\ \prod^k_{m=1}
\left (\sum^n_{i^{(m)}_0,i^{(m)}_1,\ldots i^{(m)}_{p_n-1}=1}\left (
\prod^{p^{(m)}_n}_{r=1}\xi_{i^{(m)}_{r-1}i^{(m)}_r}-E\ \prod^{p^{(m)}_n}_{r=1}
\xi_{i^{(m)}_{r-1}i^{(m)}_r}\right )\right ).
\end{split}
\end{align}
In what follows, we assume that $p^{(m)}_n$ are either $p_n$ or $p_n+1$,
$i=1,\ldots ,m$, and $p_n$ is proportional to $n^{\frac{2}{3}}$.  Let us
consider a set of $k$ closed paths
$$\p_m=\{i^{(m)}_0\rightarrow i^{(m)}_1\rightarrow\cdots\rightarrow
i^{(m)}_{p_n}=i^{(m)}_0\},\ m=1,\ldots ,k.$$  
We recall two definitions from [22], [23]:
\medskip

\noindent{\bf Definition 5}

We say that paths $\p_{m'},\ \p_{m''}$ intersect by an edge if $\p_{m'},\
\p_{m''}$ have a common (nonoriented) edge.
\medskip

\noindent{\bf Definition 6}

A subset $\p_{m_{\ell_1}},\ \p_{m_{\ell_2}},\ldots \p_{m_{\ell_k}}$
of the set of paths is called a cluster of intersecting paths if the 
following conditions hold:
\begin{enumerate}
\item[(a)] For each pair $\p_{m_i},\ \p_{m_j}$ from the subset, there exists 
a chain of paths also from the subset such that $\p_{m_i}$ is the first
path in the chain, $\p_{m_j}$ is the last path in the chain, and any two
neighboring paths intersect each other by an edge.

\item[(b)] Property (a) is violated if we add any other path from the set
to this subset.
\end{enumerate}

By definition, the sets of edges corresponding to different clusters are
disjoint.  Therefore $ E\ \prod^k_{m=1}(\sum^n_{i^{(m)}_0,
i^{(m)}_1,\ldots i^{(m)}_{p_n-1}=1}(\prod^{p^{(m)}_n}_{r=1}
\xi_{i^{(m)}_{r-1}i^{(m)}_r}-E\prod^{p^{(m)}_n}_{r=1}\xi_{i^{(m)}_{r-1}
i^{(m)}_r}))$ can be represented as a product of mathematical expectations
over disjoint clusters.  The following lemma is crucial.
\medskip

\noindent{\bf Lemma 8}
{\it
Let $p^{(1)}_n,\ldots , p^{(k)}_n\in\{2[t\cdot n^{\frac{2}{3}}],
2[t\cdot n^{\frac{2}{3}}]+1\}.$  Then 
\begin{equation}
E\ n^{-(p^{(1)}_n)+\ldots p^{(k)}_n)/2}\cdot\prod^k_{m=1}
\left (\sideset{}{^*}\sum_{i^{(m)}_0,\ldots i^{(m)}_{p^{(m)}_n-1}=1}
\left \vert \prod^{p^{(m)}_n}_{r=1}\xi_{i^{(m)}_{r-1}i^{(m)}_r}-E\left (
\prod^{p^{(m)}_n}_{r=1}\xi_{i^{(m)}_{r-1}i^{(m)}_r}\right)\right\vert\right )
\end{equation}
where the sum $\sum^*$ is over paths that form a cluster, is bounded by 
$e^{\const_k\cdot t^3}$ uniformly in $n$.  A subsum of (4.24) over clusters
in which some edges appear more than twice is going to zero.}
\medskip

To estimate the sum in (4.24) we introduced a correspondence between a set
consisting of clusters of $k$ paths and a set of even paths of length
approximately $k$ times larger.  Loosely speaking, we glue $k$ paths
together along common edges and then erase these edges.  As a result, we
get an even path of length not greater than $k\cdot p_n$ and not
smaller than $kp_n-2k$.  The details of such correspondence is
discussed in [22].  It appears that the number of preimages of an even
path under the mapping can be estimated in terms of the trajectory $X$ of
a simple random walk associated to the path.  Let $K_n(X)$ be the number
of instants $\tau_i$ of a simple random walk of length $k\cdot p_n-q$,
with $0\leq q\leq 2k$, such that $0\leq\tau_i\leq (k-1)p_n-q$ and
$x(\tau )\geq x(\tau_i)$ for $\tau_1\leq\tau\leq\tau_i+p_n$.
Then the number of preimages of $X$ is bounded by
\begin{equation}
\const_k\cdot p^{k-1}_n\cdot K_n(X)^{k-1}.
\end{equation}
The estimate in (4.25) also gives the right order.  We proved in [22] that
\begin{equation}
E_XK_n=2\cdot\sqrt{\frac{p_n}{\pi}}\cdot (1+\bar o(1)).
\end{equation}
Repeating the lines of the proof one also has
\begin{equation}
E_XK^{k-1}_n\leq \const_k\cdot p^{\frac{k-1}{2}}_n.
\end{equation}
We see that the problem is again reduced to the counting of even paths of
the length proportional $n^{\frac{2}{3}}$.  We now have a new factor const$_k
\cdot p^{k-1}_n\cdot K^{k-1}_n(X)$ in the statistical weight.
The inequality (4.27) gives us desired control of this factor and from
this point the arguments are the same as in the proof of Theorem 2.
\qed

It is left to be noted that Lemma 8 immediately implies Theorem 3.

\section{Proof of the Universality of Local Correlations at the Edge}

In \S 1 we introduced local statistics
\begin{equation}
S_{n,k}(t_1,\ldots ,t_k)=\sum_{\stackrel{j_1\neq\cdots\neq j_k;}{0\leq
\lambda_{j_1},\ldots ,\lambda_{j_k}<1+\frac{1}{2\sqrt n}}}e^{t_1\cdot
\theta_{j_1}}\cdot\ \ \ \cdot e^{t_k\cdot\theta_{j_k}}
\end{equation}
where $\theta$'s are defined by the rescaling $\lambda_j=1+\frac{\theta_j}
{2n^{\frac{2}{3}}}$ for positive eigenvalues.

By definition, it follows that $E(s_{n,k})$ is a Laplace transform of the
rescaled $k$-point correlation function restricted to the region $\theta_1,
\ldots ,\theta_k<n^{\frac{1}{6}}$.  Lemma 9 formulated in \S 1 claims that
the mathematical expectation $ES_{n,k}(t_1,\ldots ,t_k);\break t_1,\ldots ,t_k>0$,
has a universal limit as $n\rightarrow\infty$.

\noindent{\bf Proof of Lemma 9}.
Since $S_{n,k}(t_1,\ldots ,t_k)$ is a polynomial in terms of $S_{n,1}(t_1),
\ldots ,S_{n,1}(kt_1)$, $S_{n,1}(t_2)$, $\ldots ,S_{n,1}(kt_2)$, 
$\ldots ,S_{n,1}(kt_k)$, it is enough to show that the joint moments of 
$S_{n,1}(t)$ have universal limits.  But this can be recognized as an 
easy consequence of Theorem 3.  Indeed as we explained in \S 1 (see (1.28),
(1.29))
\begin{align}
\begin{split}
S_{n,1}(t)&=\sum_{\theta_j<n^{\frac{1}{6}}}e^{t\cdot\theta_j}\\
&=\frac{1}{2}
\Trace\biggl\lbrack\left (A_n^{2[t\cdot n^{\frac{2}{3}}]}+A_n^{2[t\cdot 
n^{\frac{2}{3}}]+1}\right )\cdot\chi_{I_n}(A)\biggr\rbrack \cdot
(1+\underline 0(n^{-\frac{1}{3}}))
\end{split}
\end{align}
where $\chi_{I_n}$ is an indicator of a segment $I_n=(-1+\frac{1}{2\sqrt n},\
1-\frac{1}{2\sqrt n})$.  An estimate of Chebyshev type imples that all
moments of Trace $[(A_n^{2[t\cdot n^{\frac{2}{3}}]}+
A_n^{2[t\cdot n^{\frac{2}{3}}]+1})\cdot \chi_{\bbR^1\setminus I_n} (A)]$ are
of order $0(e^{-\const\cdot n^{\frac{1}{6}}})$ and therefore negligible.
Now Corollaries 5, 6 from \S 2 finish the proof.\qed

To deduce Theorem A from Lemma 9 we note that if $\rho_{n,\beta , k}(\theta_1,
\ldots\theta_k)$ are rescaled $k$-point correlation function at the edge
for arbitrary Wigner matrix, then $\int^{+\infty}_{n^{\frac
{1}{6}}}\ldots \int^{+\infty}_{n^{\frac{1}{6}}}\rho_{n,\beta ,k}(\theta_1,
\ldots\theta_k)\, d\theta_1\ldots d\theta_k\leq \frac{n!}{(n-k)!}
e^{-\const_k\cdot n^{\frac{1}{6}}}=\bar o(1)$.  Therefore it is enough
to prove weak convergence for
\begin{equation}
\p_{n,\beta ,k}(\theta_1,\ldots\theta_k)\cdot\chi_{(-\infty
 ,n^{\frac{1}{6})^k}}
(\theta_1,\ldots \theta_k) d\theta_1\ldots d\theta_k.
\end{equation}
Multiplying (5.3) by the factor $\exp (\theta_1+\cdots +\theta_k)$ we get
a finite measure.  Since convergence of Laplace transforms of finite
measures implies weak convergence (this can be proved by Helly's theorem),
we are done.

We are now ready to prove the main result.  Let $\lambda^{(n)}_1\geq
\lambda^{(n)}_2\geq\cdots\geq\lambda^{(n)}_k$ be the first $k$ largest
eigenvalues and $s_1\geq\cdots\geq s_k$ --- an arbitrary ordered set of
real numbers.  We want to show that $P_n\{\lambda^{(n)}_1\leq 1+\frac{s_1}
{2n^{\frac{2}{3}}},\ldots ,\lambda^{(n)}_k\leq 1+\frac{s_k}
{2n^{\frac{2}{3}}}\}$ has a universal limit (Tracy-Widom distribution)
as $n\rightarrow\infty$.  In terms of $\theta$'s, the event can be written
as $\{\theta^{(n)}_1\leq s_1,\ldots ,\theta^{(n)}_k\leq s_k\}$, and
its probability can be written as a finite linear combination of
probabilities
\begin{equation}
P_n\{\#\{\theta^{(n)}_j\in (s_i,s_{i-1}],j=1,2,\ldots\}=m_i\},
i=1,\ldots ,k,
\end{equation}
$\sum^k_{i=1}m_i\leq k$.  Let us introduce $n^{(n)}_i=\#\{
\theta^{(n)}_j\in (s_i,s_{i-1}],j=1,2\ldots\}$.  By definition of correlation
functions, the factorial moments can be written as
\begin{equation}
E\prod^k_{i=1}\prod^{\ell_i-1}_{\ell =0}(\eta^{(n)}_i-\ell )=\int\rho_{n,
\beta ,L}(\theta_1,\ldots ,\theta_L)\, d\theta_1\ldots d\theta_L,
\end{equation}
where $ L=\sum^k_{i=1}\ell_i$, and integration is over
$ \prod^k_{i=1}(s_i,s_{i-1}]^{\ell_i}$.  Since correlation
functions weakly converge we deduce from (5.5) that the joint moments
of $\eta^{(n)}_i$, $i=1,\ldots k$, have universal limits as $n\rightarrow
\infty$.  It is well known that if the limiting moments grow up not faster
than factorials, they uniquely determine the limiting distribution (5.4).
The determinantal (Pfaffian) form of limiting correlation functions asserts
that this is exactly the case in our situation.  The main result is proven.
\def\am{{\it Ann. of Math.} }
\def\ap{{\it Ann. Probab.} }
\def\temf{{\it Teor. Mat. Fiz.} }
\def\jmp{{\it J. Math. Phys.} }
\def\cmp{{\it Commun. Math. Phys.} }
\def\jsp{{\it J. Stat. Phys.} }

\end{document}